%% file: main.tex
\documentclass{article}
\input{config}

\begin{document}

\twocolumn[
\icmltitle{Prompt-guided Precise Audio Editing with Diffusion Models}




\begin{icmlauthorlist}
\icmlauthor{Manjie Xu}{bit,labp}
\icmlauthor{Chenxing Li}{labp}
\icmlauthor{Duzhen zhang}{labp}
\icmlauthor{Dan Su}{labp}
\icmlauthor{Wei Liang}{bit}
\icmlauthor{Dong Yu}{labs}
\end{icmlauthorlist}

\icmlaffiliation{labp}{Tencent AI Lab Beijing}
\icmlaffiliation{labs}{Tencent AI Lab Seattle}
\icmlaffiliation{bit}{Beijing Institute of Technology}

\icmlcorrespondingauthor{Chenxing Li}{lichenxing007@gmail.com}
\icmlcorrespondingauthor{Wei Liang}{liangwei@bit.edu.cn}
\icmlcorrespondingauthor{Dong Yu}{dongyu@ieee.org}

\icmlkeywords{Audio editing, Audio generation, Diffusion}

\vskip 0.3in ]



\printAffiliationsAndNotice{} % otherwise use the standard text.

\begin{abstract}

    Audio editing involves the arbitrary manipulation of audio content through precise control. Although text-guided diffusion models have made significant advancements in text-to-audio generation, they still face challenges in finding a flexible and precise way to modify target events within an audio track. We present a novel approach, referred to as \acf{method}, which serves as a general module for diffusion models and enables precise audio editing. The editing is based on the input textual prompt only and is entirely training-free. We exploit the cross-attention maps of diffusion models to facilitate accurate local editing and employ a hierarchical local-global pipeline to ensure a smoother editing process. Experimental results highlight the effectiveness of our method in various editing tasks.

\end{abstract}

\setstretch{0.99}
\section{Introduction}

Recent progress in image synthesis~\cite{ramesh2021zero,rombach2022high,ramesh2022hierarchical} has inspired the application of text-guided diffusion models in \ac{tta} generation, known for their realism and diversity~\cite{huang2023make,ghosal2023text,liu2023audioldm,liu2023audioldm2,huang2023make2, yang2023uniaudio}. Thanks to large-scale training data and prompt-enhanced methods, these diffusion models have demonstrated great potential in modeling long continuous signal data, and have successfully learned to produce sounds based on given text. 

\begin{figure}[!t]
\centering
    \begin{subfigure}[b]{\linewidth}
    \centering
    \includegraphics[width=\linewidth]{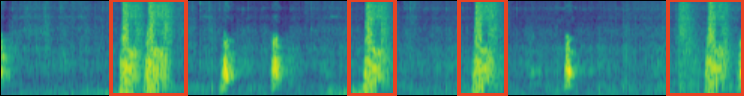}
    \caption{Source Audio: A dog barking}
    \end{subfigure}
    \begin{subfigure}[b]{\linewidth}
    \centering
    \includegraphics[width=\linewidth]{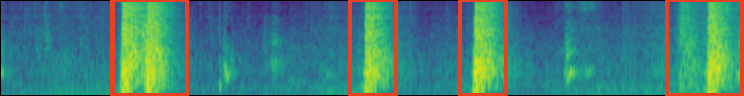}
    \caption{Precise Editing: Replace \textcolor{red}{dog barking} to \textcolor{red}{gun shooting}}
    \label{fig:intro:b}
    \end{subfigure}
    \begin{subfigure}[b]{\linewidth}
    \centering
    \includegraphics[width=\linewidth]{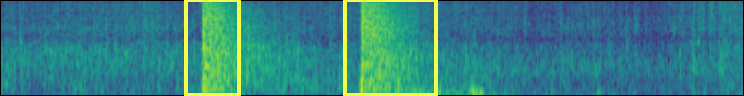}
    \caption{Traditional Editing: Replace \textcolor{red}{dog barking} to \textcolor{red}{gun shooting}}
    \label{fig:intro:c}
    \end{subfigure}
\caption{\textbf{Precise audio editing.} Such editing requires modifying the target events while preserving the unrelated events and keeping the overall structure unchanged.}
\label{fig:intro_demo}
\vspace{-10pt}
\end{figure}

However, these generative models still face challenges in editing tasks, particularly in \textbf{precise} editing. Precise audio editing involves modifying the target events within an audio track while preserving the unrelated part unchanged. As illustrated in \cref{fig:intro_demo}, precise editing (\cref{fig:intro:b}) replaces "dog barking" with "gun shooting" in the original place. In contrast, traditional editing often overemphasizes the replacement of the content itself, leveraging regeneration steps to ensure the emergence of gun shooting, while often changing the overall structure of the original audio (\cref{fig:intro:c}).

The key to precise editing lies in the ability to accurately differentiate between targeted sections for editing and unrelated parts, ensuring that manipulations are strictly confined to the intended areas. Image editing methods often achieve this by providing spatial localization masks, which can be time-consuming and labor-intensive. Furthermore, such methods cannot be easily generalized to audio editing, as target events mixed in an audio piece are often difficult to identify and separate manually. More recently, researchers mainly utilize pre-trained \ac{tta} generation models or concentrate on end-to-end training with human-provided instructions~\cite{wang2023audit,liu2023audioldm,huang2023make, yang2023usee}, while these methods do not guarantee precision in the process of regeneration. They can also be resource-intensive during the end-to-end training process, as they require a large number of editing demonstration pairs as training data.

\begin{figure*}[t!]
    \centering
    \small
    \includegraphics[width=\linewidth]{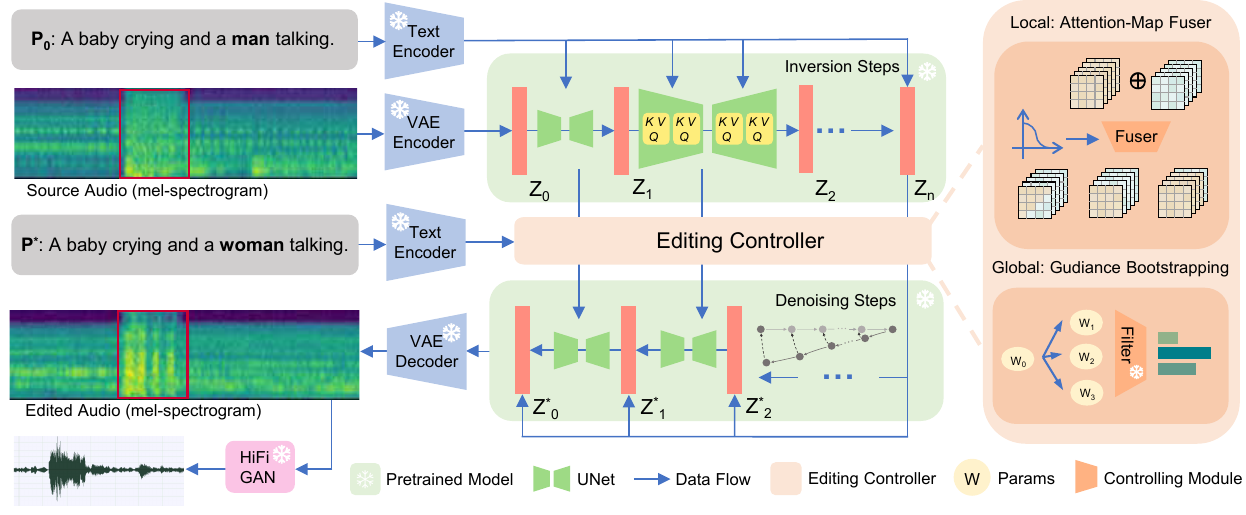}
    \caption{\textbf{The overview of the proposed \ac{method}.} Given the edit instruction, the source audio will first be inverted into the given diffusion model's domain, and then edited on the attention-map level under the guidance of our editing controller. The controller accomplishes precise editing by utilizing hierarchical guidance throughout the diffusion process. The whole editing pipeline is training-free and is adaptable to common diffusion models.}
    \label{fig:env}
\end{figure*}

In this work, we propose \acf{method}, a training-free approach for precise audio editing. Building on the success of image editing techniques that rely on attention map manipulation~\cite{hertz2022prompt,parmar2023zero,patashnik2023localizing}, \ac{method} focuses on the cross-attention layer of diffusion models, where text and audio features are interconnected. We demonstrate that diverse types of precise audio editing can be achieved by manipulating the cross-attention map during the denoising process.  Our approach serves as a flexible audio editing interface, where users only need to provide edited textual prompts. Additionally, it can function as a plug-in editing module in various diffusion models.

To perform editing, we first convert the original audio into edit-friendly noise spaces and then perform editing by injecting cross-attention maps into the diffusion process. We additionally design a hierarchical pipeline to guarantee the editing effectiveness. Locally, we import a particular Fuser module to integrate different attention maps seamlessly in single diffusion steps. This integration is crucial for mitigating the abrupt transitions often caused by sudden changes in the attention map. Globally, we employ a bootstrapping method to adjust the guidance scale, acknowledging the variability of editing targets across different audio samples. We find such a method allowing for tailored editing, adapting to the unique requirements of each audio piece. Our ablation studies confirm the effectiveness of these innovations. Moreover, we observe that the editing targets in audio differ markedly from those in images, necessitating distinct hyperparameters for optimal performance. This distinction underlines the unique challenges inherent in audio editing, and our method's adaptability in addressing these challenges. 

To the best of our knowledge, our work showcases the first attempt at utilizing attention map-level manipulation to achieve precise editing for audio. The proposed \ac{method} approach, compared with traditional audio editing methods, offers several key benefits: it achieves \textbf{precise} editing, which can ensure the manipulations are confined to the intended event; the editing process is \textbf{flexible} based on the textual prompts; it is entirely \textbf{training-free}; and it is \textbf{compatible} with widely-used diffusion models. We hope that this work serves as a step towards addressing the distinct challenges associated with audio editing\footnote{See the project page at \url{https://sites.google.com/view/icml24-ppae}.}.

\section{Related Work}
\subsection{Diffusion-based TTA Models}
Diffusion models have emerged as a promising approach for generating high-quality and diverse samples, from image synthesis~\cite{ramesh2021zero,rombach2022high,ramesh2022hierarchical} to text-to-audio generation~\cite{kreuk2022audiogen,huang2023make,ghosal2023text,liu2023audioldm}. More recently, methods like Diffsound~\cite{yang2023diffsound} and AudioLDM~\cite{liu2023audioldm,liu2023audioldm2} work based on the diffusion model to produce better \ac{tts} and \ac{tta} generation results; Tango~\cite{ghosal2023text} adopts an instruction-tuned \acf{llm} FLAN-T5 to utilizing its powerful representational ability in \acf{tta} generation; Make-An-Audio~\cite{huang2023make,huang2023make2} leverages prompt-enhanced methods to train diffusion models with high-quality text-audio pairs.
\subsection{Diffusion-based Image Editing}
Text-conditioned editing based on diffusion models has recently garnered significant interest in the image domain~\cite{shi2023dragdiffusion,kawar2023imagic,hertz2022prompt,zhang2023adding}. When compared to \ac{gan}-based methods such as DragGAN~\cite{pan2023drag}, diffusion-based approaches are considered to have better generality and higher-quality editing effects due to the advantages of diffusion models~\cite{shi2023dragdiffusion}. On the one hand, techniques like Glide~\cite{nichol2021glide} and Diffusionclip~\cite{kim2022diffusionclip} excel at global editing or detailed local editing when precise masks of the restricted area are provided. On the other hand, works such as \ac{ptp}~\cite{hertz2022prompt} achieve intuitive image editing by examining the semantic strength in cross-attention maps and performing injection. Various inversion methods~\cite{mokady2023null, huberman2023edit} have been introduced to help invert the given image, thereby facilitating further editing.
\subsection{Audio Editing}
Traditional audio editing methods predominantly focus on global editing tasks, such as audio super-resolution~\cite{birnbaum2019temporal}, audio inpainting~\cite{Adler2011audio,moliner2023diffusion,wang2023audit} and style transfer~\cite{Grinstein2018Audio,lu2019play,cifka2020groove2groove,netzorg2023permod}. Among these, uSee~\cite{yang2023usee} proposes a unified model to perform speech editing given text description and specific arguments; Loop Copilot~\cite{zhang2023loop} and InstructME~\cite{han2023instructme} enable generating and refining generated music. Recent research has started to explore more fine-grained audio editing techniques, concentrating on tasks such as adding, removing, or replacing particular audio events within a specific audio piece. Audit~\cite{wang2023audit} trains a latent diffusion model on editing tasks and supports instruction-guided audio editing. More recently, \ac{tta} generation is combined with personalization methods to meet user preferences~\cite{plitsis2024investigating}.
\subsection{Comparison}
Achieving precise editing in audio is challenging, especially when compared to visual media editing, due to the inherent temporal and spectral intricacies of audio signals. Conventional approaches like Audit and PerMod depend on training a latent diffusion model specifically for audio editing tasks and regeneration. Conversely, \ac{method} enables a training-free and flexible manipulation of audio content and is not limited to music or speech editing only. Additionally, \ac{method} offers a higher degree of granularity, empowering users to edit specific audio elements within a track—a level of precision that has been challenging to attain with existing methods. This fine-grained control also parallels the progress seen in text-conditioned image editing using diffusion models, while uniquely adapted to the audio domain in editing tasks.
\section{Method}
Formally, let $A$ represent a given audio piece, and $P$ denotes the textual description (prompt) of that audio piece; we aim to edit the input audio guided solely by the edited text prompt $P^{*}$, resulting in the final edited audio $A^{*}$. Researchers are also interested in cases where the original text prompt is absent but only with a command such as "Replace the $A$ with a $B$." We note that this scenario can be addressed through audio captioning. In this paper, we mainly focus on the former task.
\subsection{System Overview}
Our method comprises three parts: an inversion module that maps the given audio piece into the pre-trained \ac{ldm}'s domain, an \ac{ldm} pre-trained on audio, and a hierarchical editing controller which is plugged into the \ac{ldm} that facilitates audio editing. \ac{ldm}s for audio generation often contain an \ac{vae} that projects the input mel-spectrograms into the latent space, a textual-prompt encoder that transforms the text prompt into embeddings, and a diffusion network. The system's overview is illustrated in \ref{fig:env}.
\subsection{LDM}
The fundamental concept of diffusion models revolves around the iterative refinement of a randomly sampled noise input $x_{t} \sim \mathcal{N}(0, I)$, in a controlled manner, to $x_0$. With a trained perceptual compression model, \ac{ldm}s focus on the efficient, low-dimensional latent space, where the goal is to derive $z_0$ from $z_t$. With a given text description $P$, to perform sequential denoising, we train a network $\epsilon_\theta$ predict artificial noise, following the objective:
\begin{equation}
\min_\theta \mathbb{E}_{z_0, \epsilon \sim \mathcal{N}(0, I), t \sim \text{Uniform}(1, T)} \lVert \epsilon - \epsilon_\theta(z_t, t, \psi(P)) \rVert_2^2,
\end{equation}
where the condition $C = \psi(P)$ is the text embedding of the description. In \ac{tta} generation, $z$ is the latent representation of the mel-spectrogram of the audio, where researchers often leverage a pre-trained \ac{vae} to help compress the mel-spectrogram into the latent space. 

For the conditional generation of \ac{ldm}s, classifier-free guidance has proven to be an effective method for text-guided generation and editing~\cite{ho2022classifier}. When given a latent and a textual prompt, the generation is performed both conditionally and unconditionally and then extrapolated according to a given weight. Formally, let $\varnothing = \psi(\text{" "})$ be the null text embedding, the generation can be defined by:
\begin{equation}
\tilde{\epsilon}_\theta = w \cdot \epsilon_\theta(z_t, t, \psi(P)) + (1 - w) \cdot \epsilon_\theta(z_t, t, \varnothing),
\end{equation}
where $w$ denotes the guidance scale.

\subsection{Attention Map Editing}
Popular diffusion-based generation models utilize U-Nets~\cite{ronneberger2015u} with the cross-attention mechanism~\cite{vaswani2017attention} as the diffusion network. When doing conditional generation, the embeddings of different modalities are often fused in the cross-attention layers. Researchers have shown that injecting the cross-attention maps of the input enables precise editing while maintaining the original composition and structure of the original input~\cite{hertz2022prompt}. As commonly defined, we note attention maps as:
\begin{equation}
M = Softmax(\frac{QK^T}{\sqrt{d}}),
\end{equation}
where $Q = \ell_Q (\phi(z_t))$ is the query matrix of the deep spatial features of the noisy input, $K = \ell_K (\psi(P))$ and $V = \ell_V (\psi(P))$ are the key matrix and the value matrix of the textual embedding, and $\ell_Q$, $\ell_K$, $\ell_V$ are the learned linear projections.

In image editing, researchers perform replacements to get a new attention map $M^*$ when generating the new target $z_t^*$ from $P*$, and override the original attention map $M$ in the computation of a single step $t$ of the diffusion process, noted as $DM(z_t^*, P^*, t, s)\{M \leftarrow \hat{M_c}\}$. Although this approach is also applicable in \ac{tta} editing, we observe that a significant abrupt change in the diffusion step may result in the generation of indistinct audio with low quality. An intuitive solution is to incrementally incorporate the editing component into the original attention map, transitioning from a low to a high ratio. We propose a fusion mechanism, utilizing a cosine scheduler to manage the transition of the attention map during the editing process. We denote this as $M_{edit} = Fuser(M_t, M_t^*, t)$. For different editing tasks, we denote our method as follows:
\begin{equation}
Fuser := 
\begin{cases}
S_{ca}(t) \cdot M_t^* + (1 - S_{ca}(t))\cdot M_t, \\
 \qquad\qquad\qquad \text{for Audio Replace} \\
S_{ca}(t) \cdot (M_t^*)_{i,j} + (1 - S_{ca}(t))\cdot M_t, \\
 \qquad\qquad\qquad \text{for Audio Refine} \\
c \cdot S_{ca}(t) \cdot (M_t^*)_{i,j} + (1 - S_{ca}(t)) \cdot M_t, \\
 \qquad\qquad\qquad \text{for Audio Reweight} \\
\end{cases}
\end{equation}
where $S_{ca}(t)$ is the fusion ratio determined by the CosineAnnealing scheduler at step $t$, $(M_t^*)_{i,j}$ means that we only modify the pixel value $i$ according to the selected text token $j$, and $c$ denotes the scale extent of the reweighted token in Reweight task. The CosineAnnealing scheduler can be expressed as:
\begin{equation}
S_{ca}(t) = \eta_{\text{min}} + \frac{1}{2} (\eta_{\text{max}} - \eta_{\text{min}}) \left(1 + \cos\left(\pi \cdot \frac{t - t_s}{t_e - t_s}\right)\right),
\end{equation}

where $t_s$ and $t_e$ represent the starting and ending steps of the transitional phase during diffusion steps respectively, $\eta_{\text{min}}$ and $\eta_{\text{max}}$ are the minimum and maximum values for the ratio, which commonly be 1 and -1.

\subsection{Inversion}
Text-guided editing with the method mentioned above requires inverting the given audio and
textual prompt. Many previous works on text-to-image generation have focused on \ac{ddim} inversion~\cite{song2020denoising,dhariwal2021diffusion}, as \ac{ddim} sampling is considered as a deterministic sampling process that maps the initial noise to an output. 

However, such inversion has been found lacking when classifier-free guidance is applied. To overcome this issue, Null-text Inversion~\cite{mokady2023null} imports pivotal inversion for diffusion models and null-text optimization to achieve high-fidelity editing of natural images. 
DDPM-based inversion methods have also been developed~\cite{wu2022unifying,huberman2023edit}. We have adapted different inversion modules according to \ac{ldm}s, see \cref{sec:supp:models} for details.

\subsection{Guidance Bootstrapping}
The guidance scale $w$ plays a crucial role in controlling the level of importance assigned to a given prompt during the generation process in diffusion models. Generally, there exists a common scale for $w$ that enables the model to produce creative or precise outputs. For example, Stable Diffusion~\cite{rombach2022high}'s $w$ lies between 5 and 15. However, determining a universally applicable $w$ for audio generation presents a challenge, as the editing components in audio generation can vary significantly, ranging from the resounding nuances of human voices to blurred background sounds. Existing work~\cite{liu2023audioldm} has also shown the effectiveness of guidance scale on key metrics like \ac{kl} and \ac{fd}. We introduce a bootstrapping approach that aims to avoid the selection of the guidance scale. We initialize a list of $W=[w_1, w_2, w_3, ... w_n]$ before editing, and then computing $z_0$ for each $w_i$ separately. We employ a Filter module to help get the final editing results. By default, \ac{method} uses a \ac{clap} model~\cite{elizalde2023clap} as the naive filter function $f$ to identify the output with the highest relevance to the target prompt. The process can be fully paralleled. 

\subsection{Overall Framework}
The overall algorithm of \ac{method} can be described as follows: given an audio piece and its prompt, we firstly set the guidance scale to 0 and perform an inversion on the hidden space with the source prompt; then we do denoising process on the source prompt and target prompt separately, under the guidance of the \ac{method} editing controller. Locally, we edit the attention map based on the editing instructions under the guidance of Fuser, and decode the final latent variable and filter on the guidance scale to obtain the edited audio. More details can be found in \cref{sec:supp:models}.
\vspace{-5pt}
\begin{algorithm}
\label{alg:ppae}
\caption{\ac{method}}
\begin{algorithmic}[1]
\STATE \textbf{Input:} A piece of audio $a$ and its description as the source prompt $P$, a target prompt $P^*$
\STATE \textbf{[Optional for $Fuser$]:} $\eta_{\text{min}}$ and $\eta_{\text{max}}$, $t_s$ and $t_e$;\\ 
\STATE \textbf{[Optional for $Bootstrapping$]:} $W = \{w_i\}$\\
\STATE \textbf{[Optional for $Filter$]:} $Filter_w = filter_{w} (a, f)$\\
\STATE \textbf{Output:} A piece of edited audio $a^*$.
\vspace{1mm} \hrule \vspace{1mm}
\STATE $ w = 0, z_0 = Encoder(a)$
\STATE $\{z_T, z_{T-1}, ..., z_1\} = Inversion(z_0, P, w)$
\STATE $z^*_T \leftarrow z_T$;
\FOR{$t = T, T - 1, \ldots, 1$}
    \STATE $z_{t-1}, M_t \leftarrow DM(z_t, P, t)$;
    \STATE $M^*_t \leftarrow DM(z^*_t, P^*, t)$;
    \STATE $M_{ct} \leftarrow Fuser(M_t, M^*_t, t)$;
    \STATE $z^*_{t-1} \leftarrow DM(z^*_t, P^*, t, W)$ with $M \leftarrow M_{ct}$;
\ENDFOR
\STATE $a^* = Filter_{w}(Decoder(z_0^*))$
\STATE {\bfseries Return:} $a^{*}$

\end{algorithmic}
\end{algorithm}
\vspace{-10pt}
\section{Experiments}
\subsection{Editing Task}
We mainly focus on the three following audio editing tasks:

\textbf{Audio Replace: }This task aims to replace a specific audio event in a given audio piece with another, while keeping the remaining part unchanged. For example, given an audio piece "A cat meowing and then a baby crying," the task could involve changing "cat meowing" to "dog barking."

\textbf{Audio Refine: }Audio Refine here involves modifying an existing audio piece to meet additional requirements or preferences. The task aims to transform the original audio piece according to various extra adjective descriptions, such as altering the style or incorporating new features, while maintaining the essence of the music.

\textbf{Audio Reweight: }Audio Reweight is to alter the audio balance to emphasize or de-emphasize some aspects without compromising the audio's overall clarity. This could involve amplifying the sound of raindrops in a track where rain and thunder are present, or reducing the volume of background music in a conversation.

We only focused on these three local editing tasks here, as these tasks allow for the addition, drop, or replacement of certain audio elements or adjusting the balance of different sounds. These tasks can often be predictable and repetitive across different audio pieces and files.

\subsection{Test Set}
We construct our test set utilizing a cleaned subset of the Fsd50k dataset~\cite{fonseca2021fsd50k,liimage}. A pivotal aspect of precise audio editing is implementing precise modifications while maintaining the other elements of the audio unchanged. In each task, we select two distinct audio clips, treating one as the target for editing. For each task, we randomly sample 100 editing tasks as the test set. See \cref{sec:supp:testset} for the detailed construction process. We also select preliminary editing tasks as case studies.

\subsection{Metrics}
For objective metrics, we leverage commonly used metrics to evaluate the editing effects. We leverage \acf{fd}, \acf{fad}, \acf{sd}, and \acf{kl} divergence to measure the distance between the edited audio and the ground truth. Specifically, we also use \ac{clap} Score in some cases as an extra metric to calculate how well the target prompt aligns with the edited audio, as for tasks like Refine and Reweight, it is challenging to construct a corresponding target audio that can serve as ground truth for comparison. For subjective metrics, we primarily employed two metrics: Relevance, which measures how well the output audio matches the input editing prompt, and Consistency, which assesses the extent to which the original audio is edited in accordance with the editing goal. Details of these metrics can be found in \cref{sec:sub:obj} and \cref{sec:supp:sub}.

\subsection{Experimental Settings}

In this work, we primarily utilize Tango~\cite{ghosal2023text} as our \ac{tta} backbone model due to its success in \ac{tta} generation, while it's worth mentioning that our methods can be applied to a wide range of popular diffusion models. We run our experiments with 100 inference steps and retain the original hyperparameters from Tango. For editing, we run the denoising steps with 0.8 cross-replace steps, 0.0 self-replace steps, and 50 skip steps. The bootstrapping num $n$ is set to 5. We reset our Fuser configs to fit these settings, mainly $\eta_{\text{min}}$ and $\eta_{\text{max}}$, $t_s$, and $t_e$. We also leverage our reproduced \ac{ptp} for audio as an editing baseline. See \cref{sec:supp:models} for detailed discussion about the backbone model choice and baseline implementation.

\section{Results}
\label{sec:results}
\subsection{Audio Replace}

\begin{figure}[ht]
\centering
    \begin{subfigure}[b]{\linewidth}
    \centering
    \includegraphics[width=\linewidth]{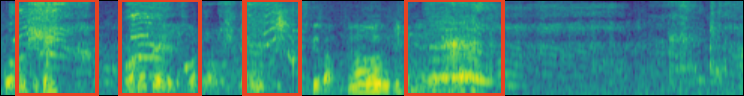}
    \caption{\label{fig:image1} A baby crying while a \textcolor{red}{man} talking.wav (Source)}
    \end{subfigure}
    \begin{subfigure}[b]{\linewidth}
    \centering
    \includegraphics[width=\linewidth]{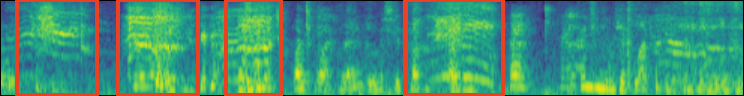}
    \caption{\label{fig:image2} A baby crying while a \textcolor{red}{woman} talking.wav (Edited)}
    \end{subfigure}
    \begin{subfigure}[b]{\linewidth}
    \centering
    \includegraphics[width=\linewidth]{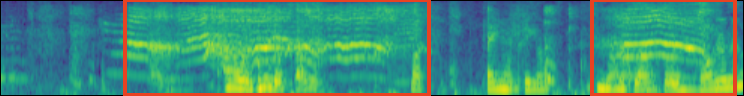}
    \caption{\label{fig:image3} A baby crying while a \textcolor{red}{woman} talking.wav (Regenerated)}
    \end{subfigure}
\caption{\textbf{Case Study (Audio Replace)}}
\label{fig:replace_demo}
\end{figure}

The \ac{method} efficiently performs replacement within a given audio piece by manipulating the attention map, and exhibits sufficient precision to preserve the overall audio structure. As demonstrated in \cref{fig:replace_demo}, the \ac{method} replaces the "man talking" event with "woman talking" while maintaining the original "baby crying" content and even the talking component in the initial audio piece, in contrast to the regenerated result. We quantitatively evaluate the replacement editing outcomes as presented in \cref{tab:replace_results}. The results reveal that the \ac{method} achieves substantial editing enhancements across the majority of metrics.

\begin{table}[!ht]
\small
\caption{\textbf{Replace Editing Results} }
\centering
\resizebox{\linewidth}{!}{
\setlength{\tabcolsep}{3mm}
\begin{tabular}{cccccc}\toprule
              \textbf{Replace} & \textbf{FAD} {\color{highlightgreen} $\downarrow$}  & \textbf{LSD} {\color{highlightgreen} $\downarrow$} & \textbf{FD} {\color{highlightgreen} $\downarrow$} & \textbf{KL} {\color{highlightgreen} $\downarrow$} & \textbf{CLAP}{\color{highlightred} $\uparrow$}\\\midrule
\ac{method} & 2.15 & \textbf{1.51} & \textbf{27.53} & \textbf{1.30} & 0.62\\\midrule
Regenerated & 4.93 & 1.74 & 32.94 & 1.69 & 0.63\\\midrule
Unedited & 1.86 & 5.98 & 45.99 & 3.28 & 0.12\\\midrule
PTP & 2.95 & 2.83 & 45.91 & 4.42 & 0.57 \\\bottomrule

   \end{tabular}}
\label{tab:replace_results}
\vspace{-5pt}
\end{table}

\begin{figure}[ht]
\centering
    \begin{subfigure}[b]{\linewidth}
    \centering
    \includegraphics[width=\linewidth]{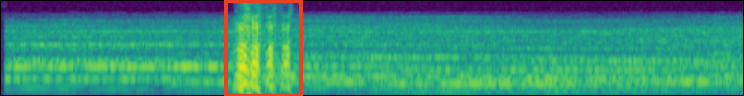}
    \caption{\label{fig:audit1} Task: Replace \textcolor{red}{laughter} to \textcolor{red}{trumpet} (Source Audio)}
    \end{subfigure}
    \begin{subfigure}[b]{\linewidth}
    \centering
    \includegraphics[width=\linewidth]{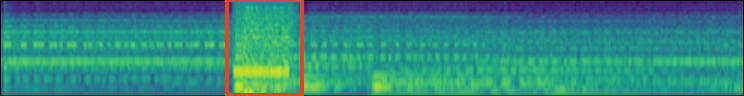}
    \caption{\label{fig:audit3} Edited by \ac{method}}
    \end{subfigure}
    \begin{subfigure}[b]{\linewidth}
    \centering
    \includegraphics[width=\linewidth]{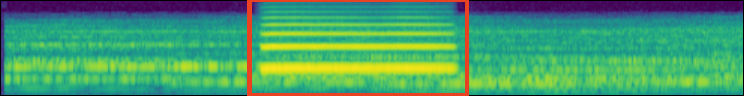}
    \caption{\label{fig:audit2} Edited by Audit}
    \end{subfigure}
    
\caption{\textbf{Case Study (\ac{method} compared with Audit)}}
\label{fig:audit_demo}
\end{figure}
 % \vspace{-20pt}
Despite the scarcity of competitive open-source baselines for audio editing tasks, we attempt to compare the \ac{method} with Audit. This alternative audio editing technique attains state-of-the-art performance in analogous tasks. The editing approach employed by the \ac{method} diverges significantly from Audit's, as the latter trains a specialized end-to-end diffusion model on editing instructions and an extensive set of data pairs. Conversely, the \ac{method} conducts training-free edits on attention map layers, rendering it compatible with a diverse range of diffusion models. Since Audit is not open-source, we compare these two audio editing techniques using Audit's publicly accessible demos, as depicted in \cref{fig:audit_demo}. For the task of "Replace laughter to trumpet," Audit regenerates the audio based on the given instruction, resulting in a change in the audio structure. On the other hand, \ac{method} only performs replacement on attention maps related to "laughter," thus preserving the original structure. The source and target prompts are generated through audio captioning and human relabeling. We want to additionally note that the editing quality of \ac{method} can be affected by the data bias between the input audio and the training set of the utilized diffusion model.

\subsection{Audio Refine}

\begin{figure}[!ht]
\centering
    \begin{subfigure}[b]{\linewidth}
    \centering
    \includegraphics[width=\linewidth]{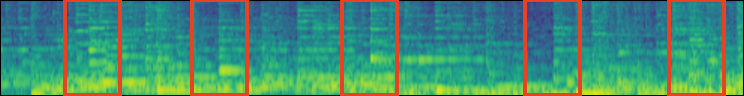}
    \caption{\label{fig:refine1} A piece of music.wav (Source)}
    \end{subfigure}
    \begin{subfigure}[b]{\linewidth}
    \centering
    \includegraphics[width=\linewidth]{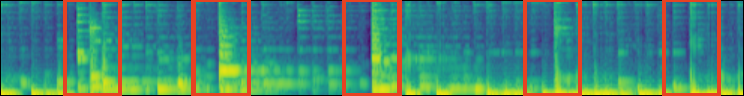}
    \caption{\label{fig:refine2} A piece of \textcolor{red}{jazz} music.wav (Edited)}
    \end{subfigure}
    \begin{subfigure}[b]{\linewidth}
    \centering
    \includegraphics[width=\linewidth]{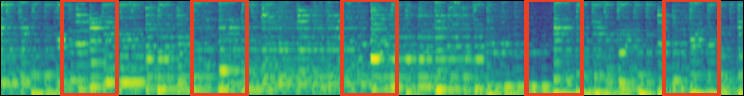}
    \caption{\label{fig:refine3} A piece of \textcolor{red}{shrill} music.wav (Edited)}
    \end{subfigure}
\caption{\textbf{Case Study (Audio Refine)}}
\label{fig:refine_demo}
\end{figure}

\ac{method} demonstrates that injecting into the attention map can aid in enhancing or modifying a given audio clip to meet supplementary requirements or preferences. As illustrated in \cref{fig:refine_demo}, when provided with the source audio "a piece of music," \ac{method} successfully refines it according to different adjective descriptions. It transforms the original audio into a different style, such as "jazz," or infuses it with a new characteristic, like "shrill," while striving to preserve the original musical content.

\begin{table}[!ht]
\small
\caption{\textbf{Refine Editing Results} }
\centering
\resizebox{\linewidth}{!}{
\setlength{\tabcolsep}{3mm}
\begin{tabular}{cccccc}\toprule
              \textbf{Refine} & \textbf{FAD} {\color{highlightgreen} $\downarrow$}  & \textbf{LSD} {\color{highlightgreen} $\downarrow$} & \textbf{FD} {\color{highlightgreen} $\downarrow$} & \textbf{KL} {\color{highlightgreen} $\downarrow$} & \textbf{CLAP}{\color{highlightred} $\uparrow$}  \\\midrule
\ac{method} & \textbf{6.86} & \textbf{1.55} & \textbf{43.31} & 1.92 & \textbf{0.63}\\\midrule
Unedited & 8.19 & 1.70 & 49.91 & 1.85 & 0.25\\\midrule
PTP & 9.35 & 1.94 & 45.88 & 1.17 & 0.32 \\\bottomrule

   \end{tabular}}
\label{tab:refine_results}
\vspace{-5pt}
\end{table}

We present the Refine editing effects in \cref{tab:refine_results}. For comparison purposes, we report the corresponding metrics of the \ac{method} editing results alongside the unedited audio. It is worth noting that obtaining an audio with the same structure that also satisfies the prompt is challenging. Therefore, we regenerate the audio according to the target prompt as the ground truth, which results in worse performance in terms of distance metrics. Nevertheless, the \ac{method} results still outperform the unedited audio, demonstrating significant editing effects. For the \ac{clap} score, we employ CLAP to calculate the similarity between the edited audio and the target prompts. The \ac{clap} scores are softmaxed. Generally, the results demonstrate that the refined editing effects significantly improve the audio quality and better align with the target prompts.

\subsection{Audio Reweight}
\begin{figure}[!ht]
\centering
    \begin{subfigure}[b]{\linewidth}
    \centering
    \includegraphics[width=\linewidth]{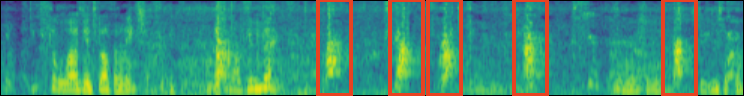}
    \caption{A woman talking and a \textcolor{red}{dog barking}.wav, $c$ = 2}
    \end{subfigure}
    \begin{subfigure}[b]{\linewidth}
    \centering
    \includegraphics[width=\linewidth]{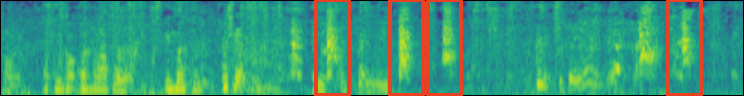}
    \caption{A woman talking and a \textcolor{red}{dog barking}.wav, $c$ = 0}
    \end{subfigure}
    \begin{subfigure}[b]{\linewidth}
    \centering
    \includegraphics[width=\linewidth]{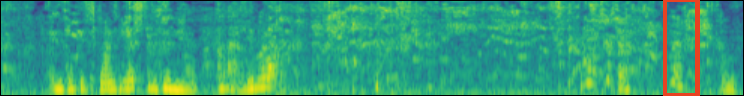}
    \caption{A woman talking and a \textcolor{red}{dog barking}.wav, $c$ = -2}
    \end{subfigure}
\caption{\textbf{Case Study (Audio Reweight), controlled token = \textit{"dog barking"}}}
\label{fig:reweight_demo}
\end{figure}

\begin{table}[!ht]
\small

\caption{\textbf{Reweight Editing Results}}
\centering
\resizebox{\linewidth}{!}{
\begin{tabular}{ccccccc}\toprule
              \textbf{Reweight} & \textbf{Original} & \textbf{$2$}      & \textbf{$1$}      & \textbf{$0$}      & \textbf{$-1$}    & \textbf{$-2$} \\\midrule
Reweight {\color{highlightred} $\uparrow$}{\color{highlightgreen} $\downarrow$}   & 0.74   & \textbf{0.83} & 0.73 & \textbf{0.35} & \textbf{0.10} & \textbf{0.12}\\\midrule
Reweight(PTP) {\color{highlightred} $\uparrow$}{\color{highlightgreen} $\downarrow$}  & 0.74 & 0.79 & 0.74 & 0.49 & 0.32 & 0.25 \\\midrule

Unrelated {\color{gray} $\rightarrow$}        & 0.91   & 0.93 & 0.91 & 0.89 & 0.82 & 0.85  \\\bottomrule
\end{tabular}}
\label{tab:reweight}
\vspace{-5pt}
\end{table}

Our methods demonstrate effective control in strengthening or weakening a specific audio event based on the textual token. \cref{fig:reweight_demo} illustrates edited audio with prompts "A woman talking and a dog barking," where we reweight on the "barking" effect. Results show that the barking component in the edited audio is controlled according to the specified controlling parameter $c$. 

We mainly employ the CLAP Score to quantitatively evaluate the reweight degree of the target event in the given audio. For the results presented in \cref{tab:reweight}, we show the CLAP scores for the reweighted event and unrelated event in reweight editing. See the error bars in \cref{tab:error_bar_tab3}. Take "A woman talking and a dog barking" as an example. If we want to reweight the "dog barking" component, we compute the CLAP score between the edited audio and "dog barking" to obtain the reweight score, and "a woman talking" to get the unrelated score. Our edited results demonstrate that a positive controlling parameter effectively strengthens the reweighted component in the audio, enabling the CLAP model to recognize it more accurately. Conversely, a negative controlling parameter assists in weakening this component. Also, the results indicate that as the controlling parameter decreases from 2 to -2, the reweighted component diminishes. The unrelated components remain relatively stable, showing that the edit will not change the other attributes of the audio.

\subsection{Subjective Evaluation}
\begin{table}[!ht]
\small
\caption{\textbf{Subjective Evaluation Results}}
\centering
\resizebox{\linewidth}{!}{
\begin{tabular}{ccccccc}\toprule
              \multirow{2}{*}{\textbf{Metric}} & \multicolumn{2}{c}{\textbf{Replace}}      & \multicolumn{2}{c}{\textbf{Refine}}      & \multicolumn{2}{c}{\textbf{Reweight}}     \\\cmidrule{2-7} 
               & \textit{\ac{method}} & \textit{Comp}      & \textit{\ac{method}} & \textit{Comp}      & \textit{\ac{method}} & \textit{Comp}    \\\midrule
Relavence {\color{highlightred} $\uparrow$}   & \textbf{95.71} & 89.28 & 81.42 & 81.42 & \textbf{99.28} & 92.14  \\\midrule
Consistency {\color{highlightred} $\uparrow$}  & \textbf{95.0} & 81.42 & \textbf{85.71} & 81.42 & \textbf{94.28} & 82.85 \\\bottomrule
\end{tabular}}
\label{tab:substudy}
\vspace{-5pt}
\end{table}

We assessed the editing effects of the \ac{method} through a Subjective Evaluation. For each task, we engaged 14 participants to evaluate the editing effect and precision on 30 randomly sampled edited audio pairs. We leverage regenerated audio based on $P^*$ as the comparison. The results in \cref{tab:substudy} show that \ac{method} significantly improves the precision of the editing while maintaining a good editing effect, whereas the regenerating baseline failed in terms of consistency. See the error bars in \cref{tab:error_bar_substudy}. More details of our Subjective Evaluation can be found in \cref{sec:supp:sub}.

\subsection{Ablation Study}

\begin{table}[!ht]
\small
\caption{\textbf{Ablation study on the generation configuration.} We show editing results with different generation configurations, especially the guidance scale $w$, the cross-replace steps $Cross$, and the self-replace steps $Self$. }
\centering
% \resizebox{\linewidth}{!}{
\begin{tabular}{cccccc}
\toprule
\textbf{Parameters} & \textbf{FAD}{\color{highlightgreen} $\downarrow$} & \textbf{LSD}{\color{highlightgreen} $\downarrow$} & \textbf{FD}{\color{highlightgreen} $\downarrow$} & \textbf{KL}{\color{highlightgreen} $\downarrow$} & \textbf{CLAP}{\color{highlightred} $\uparrow$} \\ \midrule
 7/0.8/0 & 5.97 & 2.01 & 40.90 & 1.88 & 0.51 \\\midrule
 25/0.8/0 & \textbf{3.34} & \textbf{1.62} & \textbf{25.74} & \textbf{0.62} & \textbf{0.68} \\\midrule
 75/0.8/0 & 3.67 & 1.98 & 43.24 & 1.83 & \textbf{0.69} \\\midrule
 75/0.8/0.2 & 12.92 & 1.73 & 48.90 & 1.62 & 0.57 \\\midrule
 75/0.6/0.2 & 6.79 & 2.11 & 35.63 & 1.37 & 0.41 \\ \bottomrule
\end{tabular}
% }
\vspace{-10pt}
\label{tab:config_ablation}

\end{table}

\begin{table}[!ht]
\small
\caption{\textbf{Ablation study on the Fuser with different schedulers.}  Editing results with different schedulers show that a scheduler can help generate better-edited audios. }
\centering
% \resizebox{\linewidth}{!}{
\setlength{\tabcolsep}{2.5mm}
\begin{tabular}{ccccc}
\toprule
\textbf{Fuser Scheduler} & \textbf{FAD}{\color{highlightgreen} $\downarrow$} & \textbf{LSD}{\color{highlightgreen} $\downarrow$} & \textbf{FD}{\color{highlightgreen} $\downarrow$} & \textbf{KL}{\color{highlightgreen} $\downarrow$} \\ \midrule
PTP(w/o)  & 5.36 & 2.09 & 31.32 & 0.72 \\\midrule
Exponential  & 3.65 & \textbf{1.51} & 35.02 & 1.21 \\\midrule
Linear  & 3.47 & 1.61 & \textbf{25.74} & \textbf{0.62} \\\midrule
CosineAnnealing  & \textbf{3.15} & 1.73 & \textbf{25.75} & \textbf{0.63} \\ \bottomrule
\end{tabular}
% }
\label{tab:shecduler_ablation}
\vspace{-10pt}
\end{table}

\subsubsection{Generation Configurations}
Specific generation configurations significantly impact the output quality. In particular, configurations derived from image generation and editing sometimes translate poorly when applied to audio. Among these configurations, guidance scales and replace steps are vital for successful generation and editing effects.  

We conduct several groups of editing studies, as shown in \cref{tab:config_ablation}. Generally, generation with the cross-replace steps around 0.8 and self-replace steps around 0 improves the metrics. While $w$ around 25 achieves the best performance, the generation can also potentially work with a larger or smaller guidance scale. These results also demonstrate the importance of guidance bootstrapping, as we cannot find a universally applicable $w$. A more extensive guidance scale results can lead to a more significant editing effect, as evidenced by the CLAP score, although it may affect the quality of the generated audio. An alternative solution is to increase the diffusion steps while keeping the guidance scale small. We have concluded that $steps = 1000 / w$ generally works. 

\subsubsection{Attention-Map Fuser Scheduler}
The Fuser scheduler controls when and how the attention map of the source audio and the edited audio are mixed, thus being the critical module in the proposed editing pipeline here. In \ac{method}, we use the CosineAnnealing scheduler, while here we compare the editing effects of different schedulers like linear and exponential schedulers without bootstrapping. Details of these schedulers can be found in \cref{sec:sub:sche}. The results are shown in \cref{tab:shecduler_ablation}.

We have observed that the Fuser module enhances the editing effects compared to the original \ac{ptp} model. Additionally, editing results slightly vary when different Fuser schedulers are used, with the CosineAnnealing scheduler proving slightly superior to the linear and exponential schedulers. Interestingly, the \ac{ptp} model without a Fuser exhibits the most drastic changes (\cref{fig:curve}), while the Cosine function is the most subtle, which corresponds to the editing effects. A plausible explanation could be that it facilitates a smoother combination of noisy latents, thereby aiding in merging sources from different audios. This conclusion has been similarly tested on images, where researchers linearly interpolate the noisy latent from various sources~\cite{dong2023prompt}. While they observed cluttered content in images due to the spatial (rather than semantic) mix of the source object and the edited object, we find such a problem is quite different in audio editing.

\subsubsection{Discussion about the Hyperparameters}
There are a number of hyperparameters introduced by this approach such as the number of inference steps, the guidance scale, and the parameters of the cosine scheduler. While the method is training-free in that the diffusion model need not be re-trained, results indicate that selection of these parameters is important for performance. We consider this aspect not as a drawback, but as a strategic design choice. Firstly, for the majority of editing tasks, there exist common selections for these hyperparameters that have been empirically found to perform well across a wide range of scenarios. This standard configuration serves as a solid starting point for users, ensuring that effective editing can be achieved without the need for extensive parameter tuning in general cases. For certain parameters, such as the guidance scale, we have developed a bootstrapping strategy that simplifies the selection process. This strategy assists users in automatically identifying appropriate parameter values based on the characteristics of their specific audio editing tasks, reducing the complexity involved in manual tuning. Secondly, the inclusion of these hyperparameters was a deliberate choice to provide users and researchers with the flexibility needed to fine-tune the editing process according to specific needs and constraints. This flexibility is paramount in the diverse field of audio editing, where the optimal settings for hyperparameters can significantly vary depending on the task at hand and the specific characteristics of the audio data being processed.

More details and further ablation studies on other configurations can be found in \cref{sec:supp:abl}.

\section{Audio Refusion}
The previous success of audio editing based on attention maps sets the stage for us to tackle a more challenging task called Audio Refusion: given two audio pieces, \( a_1 + a_2 \) and \( a_3 + a_4 \), the goal is to create a fusion, such as \( a_1 + a_4 \). We introduce Audio Refusion here to further demonstrate \ac{method}'s editing capability.

\begin{figure}[!ht]
\centering
    \begin{subfigure}[b]{\linewidth}
    \centering
    \includegraphics[width=\linewidth]{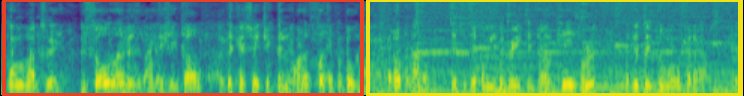}
    \caption{\textcolor{red}{A man talking} and \textcolor{arylideyellow}{a soft music}.wav (Source)}
    \end{subfigure}
    \begin{subfigure}[b]{\linewidth}
    \centering
    \includegraphics[width=\linewidth]{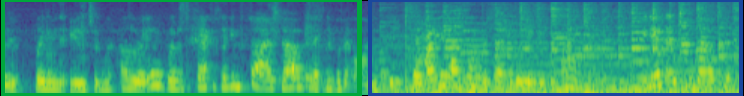}
    \caption{\textcolor{ao(english)}{A woman talking} and \textcolor{blue}{a jazz music}.wav (Source)}
    \end{subfigure}
    \begin{subfigure}[b]{\linewidth}
    \centering
    \includegraphics[width=\linewidth]{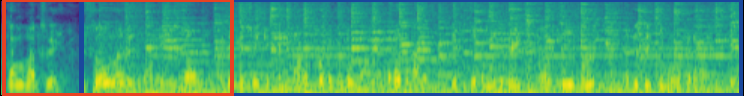}
    \caption{\textcolor{red}{A man talking} and \textcolor{blue}{a jazz music}.wav (Refused)}
    \end{subfigure}
    \begin{subfigure}[b]{\linewidth}
    \centering
    \includegraphics[width=\linewidth]{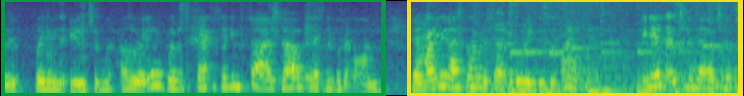}
    \caption{\textcolor{ao(english)}{A woman talking} and \textcolor{arylideyellow}{a soft music}.wav (Refused)}
    \end{subfigure}
\caption{\textbf{Case Study (Audio Refuse)}}
\label{fig:refuse_demo}
\end{figure}

We perform attention map fusion as follows:

\begin{equation}
Fuser := S_{ca}(t) \cdot (M_{1,t})_{i,j} + (1 - S_{ca}(t)) \cdot (M_{2,t})_{k, l}, 
\end{equation}

where $M_1$ and $M_2$ represent the attention maps from the two given audio sources, $(M_1)_{i,j}$ and $(M_2)_{k,l}$ denote different components of these audio sources, such as the selected text token $j$ from the first audio and the selected text token $l$ from the second audio. We fuse their attention maps $M_i$ and $M_k$ under the guidance of the fuser scheduler. It is important to note that the scheduler should aim for a more balanced fusion, as we are combining the attention maps rather than replacing them. We recommend setting $\eta_{\text{min}}$ and $\eta_{\text{max}}$ to 0.4 and 0.6, respectively, for optimal results.

Audio Fusion offers significant creative opportunities in audio production while nevertheless presenting considerable challenges. We demonstrate the fusion editing in \cref{fig:refuse_demo}. As shown, we are given two distinct audio pieces featuring human speech and different music backgrounds, and attempt to arbitrarily combine the speech with different music backgrounds through one-step editing. The \ac{method} successfully fuses audio components from different sources by utilizing inversion and fusion techniques in the attention map. Although the original audio content of the event is preserved, as shown in \cref{fig:refuse_demo}, we observe that such fusion does not guarantee precise editing at the structural level. This can be attributed to the fact that attention map fusion does not inherit the structure from the source audio, unlike injection methods used in previous tasks such as replacement.

\section{Conclusion}

In this paper, we introduce a novel approach \ac{method} for precise audio editing. By adeptly manipulating the attention map of a pre-trained diffusion model, we have demonstrated a training-free and adaptive approach for precise audio editing within diffusion models. Our approach facilitates a wide range of audio editing tasks, including content replacement and recombination, but does so while preserving the semantic essence and overall structure of the original audio. The experiments conducted showcase its potential as a highly effective editing tool. We hope our work can offer a new horizon in audio processing that is both precise and flexible to a myriad of audio editing needs.

\paragraph{future work} We see several potential avenues for enhancing the capabilities and applications of our PPAE framework. Briefly, 1) While PPAE demonstrates significant advancements in editing efficiency and flexibility, further research is needed to enhance the overall audio quality, particularly in complex editing scenarios where multiple audio elements interact. 2) Currently, PPAE and similar frameworks require processing time that limits their use in real-time applications, as the diffusion process is relatively slow. 3) Ethical and Responsible Use: As with any powerful generative technology, it's crucial to continue exploring mechanisms for ensuring the ethical use of PPAE, including safeguards against misuse for creating misleading or harmful content.

\section{Impact Statements}
The proposed \ac{method} is thought beneficial to audio generation and productions related to audio, as it provides a more precise and flexible way to edit audio content automatically. It can potentially enhance the efficiency and effectiveness of audio editing processes in various industries, and make audio editing more accessible to a broader range of users. The potential negative social impact of our methods mainly lines in the fact that precise audio editing technologies can give the evolving landscape of digital content creation and the potential for misuse in generating fake or misleading content. Also, training-free editing could make unauthorized editing of personal audio recordings easier, which could lead to privacy violations. We advocate for increased focus on addressing challenges related to the authenticity of content and its ethical utilization.

\paragraph{Acknowledgement}

We would like to thank Dr. Chi Zhang (BIGAI) and Prof. Yixin Zhu (PKU) for helpful discussions. This paper is supported by the Tencent AI Lab, the National Key R\&D Program of China (2022ZD0114900), and the NSFC (62172043).

\newpage
\appendix

\bibliography{reference}
\bibliographystyle{icml2024}

%%%%%%%%%%%%%%%%%%%%%%%%%%%%%%%%%%%%%%%%%%%%%%%%%%%%%%%%%%%%%%%%%%%%%%%%%%%%%%%
%%%%%%%%%%%%%%%%%%%%%%%%%%%%%%%%%%%%%%%%%%%%%%%%%%%%%%%%%%%%%%%%%%%%%%%%%%%%%%%
% APPENDIX
%%%%%%%%%%%%%%%%%%%%%%%%%%%%%%%%%%%%%%%%%%%%%%%%%%%%%%%%%%%%%%%%%%%%%%%%%%%%%%%
%%%%%%%%%%%%%%%%%%%%%%%%%%%%%%%%%%%%%%%%%%%%%%%%%%%%%%%%%%%%%%%%%%%%%%%%%%%%%%%
\newpage
\appendix
% \onecolumn
\input{supp}
%%%%%%%%%%%%%%%%%%%%%%%%%%%%%%%%%%%%%%%%%%%%%%%%%%%%%%%%%%%%%%%%%%%%%%%%%%%%%%%
%%%%%%%%%%%%%%%%%%%%%%%%%%%%%%%%%%%%%%%%%%%%%%%%%%%%%%%%%%%%%%%%%%%%%%%%%%%%%%%

\end{document}

%% file: config.tex
\usepackage[accepted]{icml2024}

\usepackage{times}
\usepackage{soul}
\usepackage{url}
\usepackage[hidelinks]{hyperref}
\usepackage[utf8]{inputenc}
\usepackage{caption}
\usepackage{graphicx}
\usepackage{booktabs}
\usepackage{amsmath}
\usepackage{amsthm}
\usepackage{booktabs}
\usepackage{algorithm}
\usepackage[printonlyused]{acronym}
\usepackage[switch]{lineno}
\usepackage{amsfonts}
\usepackage{amssymb}
\usepackage{cleveref}
\usepackage{xcolor}
\usepackage{setspace}
\usepackage{subcaption}
\usepackage{algorithmic}
\usepackage{multirow}
\theoremstyle{plain}

\theoremstyle{definition}

\theoremstyle{remark}

\definecolor{highlightgreen}{HTML}{009901}
\definecolor{highlightred}{HTML}{FD6864}
\definecolor{ao(english)}{rgb}{0.0, 0.5, 0.0}
\definecolor{arylideyellow}{rgb}{0.91, 0.84, 0.42}

\usepackage[textsize=tiny]{todonotes}

\acrodef{tta}[TTA]{text-to-audio}
\acrodef{method}[PPAE]{Prompt-guided Precise Audio Editing}
\acrodef{ptp}[PTP]{Prompt-to-Prompt}
\acrodef{llm}[LLM]{Large Language Models}
\acrodef{clap}[CLAP]{contrastive language-audio pretraining}
\acrodef{gan}[GAN]{Generative Adversarial Network}
\acrodef{ldm}[LDM]{Latent Diffusion Model}
\acrodef{vae}[VAE]{variational auto-encoder}
\acrodef{ddim}[DDIM]{Denoising Diffusion Implicit Models}
\acrodef{ddpm}[DDPM]{Denoising Diffusion Probabilistic Models}
\acrodef{fd}[FD]{Fréchet distance}
\acrodef{fad}[FAD]{Fréchet audio distance}
\acrodef{sd}[SD]{Spectral distance}
\acrodef{kl}[KL]{Kullback–Leibler}
\acrodef{tts}[TTS]{test-to-sound}

\icmltitlerunning{Prompt-guided Precise Audio Editing with Diffusion Models}

\newcommand{\norm}[1]{\left\Vert #1 \right\Vert_2}

%% file: supp.tex
\clearpage
\section{More editing results}
\subsection{Audio Replace}
\input{supp_fig_replace}
\clearpage
\subsection{Audio Refine}
\input{supp_fig_refine}
\clearpage
\subsection{Audio Reweight}
\input{supp_fig_reweight}

\clearpage
\section{\acs{tta} Models and Preliminaries}
\label{sec:supp:models}
\subsection{\ac{tta} Models}
We leverage Tango~\cite{ghosal2023text} as our baseline model in this work for its ability to understand complex concepts in the textual description. We note that our proposed method is also compatible with widely-used diffusion structures, such as Audioldm~\cite{liu2023audioldm, liu2023audioldm2} and Make-An-Audio~\cite{huang2023make, huang2023make2}. In general, these models consist of three primary components: i) a textual-prompt encoder, ii) a \ac{ldm}, and iii) a mel-spectrogram/audio \ac{vae}. The textual-prompt encoder processes the input audio description, which is then utilized to create a latent audio representation or audio prior to standard Gaussian noise through reverse diffusion. Following this, the mel-spectrogram VAE's decoder generates a mel-spectrogram from the latent audio representation. Finally, a vocoder receives the mel-spectrogram as input to produce the resulting audio. Our \ac{method} method only influences the \ac{ldm} part.

\subsection{Inversion}
We elaborate on the inversion function in \cref{alg:ppae}, which involves extracting a sequence of noise vectors that can reconstruct the given source content (image or audio) when used in the reverse diffusion process. Generally, there are two main categories of inversion studied: \ac{ddim} inversion and \ac{ddpm} inversion. The \ac{ddim} scheme employs a deterministic sampling process that maps a single initial noise vector to a generated image, making \ac{ddim} inversion relatively simpler. We implement \ac{ddim} inversion following the well-known Null-text Inversion algorithm~\cite{mokady2023null}, as depicted in \cref{alg:null}:

\begin{algorithm}[h]\label{alg}
\begin{algorithmic}
\STATE \textbf{Input:} A source prompt embedding $\mathcal{C} = \psi(\mathcal{P})$ and input image $\mathcal{I}$.\\
\STATE \textbf{Output:} Noise vector $z_T$ and optimized embeddings $\{\varnothing_t\}_{t=1}^{T}$ .\\ 
\vspace{1mm} \hrule \vspace{1mm}
\STATE Set guidance scale $w=1$; \\
\STATE Compute the intermediate results  $z^*_T,\ldots,z^*_0$ using DDIM inversion over $\mathcal{I}$; \\
\STATE Set guidance scale $w=7.5$; \\
\STATE Initialize $\bar{z_T} \leftarrow z^*_T$, $\varnothing_T \leftarrow  \psi("")$; \\
\FOR{$t=T,T-1,\ldots,1$}
    \FOR{$j=0,\ldots,N-1$}
        \STATE $\varnothing_t \leftarrow  \varnothing_t - \eta \nabla_{\varnothing} \norm{z^*_{t-1}-z_{t-1}(\bar{z_t}, \varnothing_t, \mathcal{C})}^2$;
    \ENDFOR
    \STATE Set $\bar{z}_{t-1} \leftarrow z_{t-1}(\bar{z_t}, \varnothing_t, \mathcal{C})$, $\varnothing_{t-1} \leftarrow  \varnothing_{t} $;
\ENDFOR
\STATE \textbf{Return} $\bar{z_T}$, $\{\varnothing_t\}_{t=1}^{T}$
\caption{Null-text inversion (DDIM Inversion)}
\label{alg:null}
\end{algorithmic}
\end{algorithm}

However, it has been observed that such a \ac{ddim} inversion method only becomes effective when a large number of diffusion timesteps are used. Even then, it often results in less than optimal outcomes in text-guided editing. Although the native \ac{ddpm} noise space is not conducive to editing, researchers have attempted to employ alternative inversion methods to fit better and achieve a more controllable editing space. We adopt the Edit-friendly Inversion~\cite{huberman2023edit} in \ac{ddpm} scenarios, as demonstrated in \cref{alg:example}:

\begin{algorithm}[!ht]
   \caption{Edit-friendly DDPM inversion}
   \label{alg:example}
\begin{algorithmic}
   \STATE {\bfseries Input:} real image $x_0$ 
   \STATE {\bfseries Output:} $\{x_T,z_T,\ldots,z_1\}$
    \vspace{1mm} \hrule \vspace{1mm}
   \FOR{$t=1$ {\bfseries to} $T$}
   \STATE $\tilde{\epsilon} \sim \mathcal{N}(0,\,1)$
   \STATE $x_t \leftarrow \sqrt{\bar{\alpha_t}} x_0 + \sqrt{1-\bar{\alpha_t}} \epsilon$
   \ENDFOR
   \FOR{$t=T$ {\bfseries to} $1$}
   \STATE $z_t \leftarrow (x_{t-1} -\hat\mu_t(x_{t})) / \sigma_t $   
   \STATE $x_{t-1} \leftarrow \hat\mu_t(x_{t})+\sigma_t z_t$ \quad // to avoid error accumulation
   \ENDFOR
   \STATE {\bfseries Return:} $\{x_T,z_T,\ldots,z_1\}$
\end{algorithmic}
\end{algorithm}

\subsection{Baselines}
We reimplement the \ac{ptp} method~\cite{hertz2022prompt} for audio editing as one of our baselines, primarily because it is closely related to the editing techniques we are exploring. Originally, the \ac{ptp} method was designed for image editing, where attention maps from the original image are injected into the edited image's attention maps during the diffusion process. To adapt this method for audio input, we migrate \ac{ptp} to work with \ac{tta} \ac{ldm}s, focusing on the latent representation of the mel-spectrogram for the input audio. It is worth noting that most existing editing works utilizing \ac{ptp} are based on \ac{ddim} inversion. However, for a fair comparison, we adopt the \ac{ddpm} inversion component from \ac{method} as the inversion mechanism for \ac{ptp}. 

While both \ac{method} and \ac{ptp} have migrated their editing pipelines from image to audio, they differ in two main aspects. Firstly, the core function in editing attention maps is distinct. Audio editing tasks present unique challenges, and a simple injection on attention map layers may not be effective. Secondly, \ac{tta} \ac{ldm}s perform differently during the generation process. Given the complexity involved in the generation and editing process, \ac{method} requires a global configuration scheduler to boost the quality of the generation results. In the main paper, we primarily discuss the influence of the guidance scale. However, it's worth noting that configurations like generation steps and attention replacement steps also play a crucial role. These aspects are demonstrated in various ablation studies that we have conducted.

\section{Objective Evaluation}
\label{sec:sub:obj}
\paragraph{\acl{fd}} \ac{fd} is a mathematical metric used to measure the similarity or dissimilarity between two curves or sequences in a metric space. In the context of audio, Frechet Distance can be used to compare generated audio samples with target samples.

\paragraph{\acl{fad}} Inspired by the Fréchet Inception Distance used in image processing, \ac{fad} measures the similarity between the distribution of features in the source audio and the edited audio. A lower \ac{fad} score indicates that the edited audio is closer to the original in terms of the overall distribution of its features, implying a higher fidelity of the editing process.

\paragraph{\acl{sd}} This metric evaluates the difference in the spectral characteristics between the original and edited audio. The \ac{sd} gives a quantitative measure of how much the frequency content has been altered during the editing process.

\paragraph{\acl{kl} Divergence} KL divergence measures how one probability distribution diverges from the expected one. In the context of audio editing, it is used to compare the distribution of certain audio features between the original and edited audio. 

\paragraph{\ac{clap} Score} For tasks like refine and reweight, it is challenging to construct a corresponding target audio that can serve as ground truth for comparison. Therefore, we leverage the \ac{clap} as an extra metric to calculate how well the target prompt aligns with the edited audio. The pre-trained \ac{clap} model extracts a latent representation of the given audio and text and returns the audio-text similarity score.

FAD, FD, and KL are well-established and widely accepted metrics in text-to-audio generation tasks. Previous works, such as AudioLDM~\cite{liu2023audioldm, liu2023audioldm2}, Tango~\cite{ghosal2023text}, and AudioGen~\cite{kreuk2022audiogen}, have similarly employed these metrics. Our intention is not to claim novelty in using these metrics but to adhere to common practices within the domain, ensuring that our evaluation is convincing and reliable. To compute these metrics, we follow the same evaluation pipeline as AudioLDM, utilizing code from the official repository (https://github.com/haoheliu/audioldm\_eval).

\section{Subjective Evaluation}
\label{sec:supp:sub}
\subsection{Evaluation Metrics}
The evaluation involves judging the relevance and consistency of each edited audio file with its file name and the original audio file, with scores ranging from 1 (lowest) to 100 (highest). Relevance refers to the match between the audio and the input textual prompt, whether the text content appears in the audio, and whether the audio corresponds to the text's semantics. Consistency assesses the degree of similarity between the current audio and the original audio. Note that the current audio and the original audio have different descriptions; for example, the current audio is "a man speaking and a dog barking," while the original audio is "a woman speaking and a dog barking." The consistency assessment focuses on whether the content and rhythm of the man's and woman's speech are consistent, as well as whether the dog's barking is consistent, without paying attention to the difference between "man" and "woman."
\subsection{Test Data}
The test set consists of fifteen different audio sets for three editing tasks mentioned in the paper, each containing one original audio file (xxx-0.wav) and two edited audio files (xxx-1.wav and xxx-2.wav). The evaluation process involves scoring each edited audio file based on the criteria mentioned above, with a detailed scoring breakdown provided for both relevance and consistency. The aim is to ensure a comprehensive understanding of the audio editing quality and its effectiveness in achieving the desired editing goals.
\subsection{Results with Error Bars}
\begin{table*}[h!]
\small
\caption{\textbf{Error bars of Reweight CLAP score}}
\centering
\begin{tabular}{ccccccc}
\toprule
 & \textbf{Original} & \textbf{2} & \textbf{1} & \textbf{0} & \textbf{$-1$} & \textbf{$-2$} \\\midrule
Reweight & $0.74\pm0.13$ & $0.83\pm0.09$ & $0.73\pm0.14$ & $0.35\pm0.18$ & $0.10\pm0.07$ & $0.12\pm0.06$ \\\midrule
Reweight(PTP) & $0.74\pm0.14$ & $0.79\pm0.18$ & $0.74\pm0.25$ & $0.49\pm0.23$ & $0.32\pm0.06$ & $0.25\pm0.09$ \\\midrule
Unrelated & $0.91\pm0.08$ & $0.93\pm0.11$ & $0.91\pm0.07$ & $0.89\pm0.13$ & $0.82\pm0.08$ & $0.847\pm0.13$ \\
\bottomrule
\end{tabular}
\label{tab:error_bar_tab3}
\end{table*}

\begin{table*}[!ht]
\small
\caption{\textbf{Error bars of Subjective Evaluation Results}}
\centering
\begin{tabular}{ccccccc}\toprule
              \multirow{2}{*}{\textbf{Metric}} & \multicolumn{2}{c}{\textbf{Replace}}      & \multicolumn{2}{c}{\textbf{Refine}}      & \multicolumn{2}{c}{\textbf{Reweight}}     \\\cmidrule{2-7} 
               & \textit{\ac{method}} & \textit{Comp}      & \textit{\ac{method}} & \textit{Comp}      & \textit{\ac{method}} & \textit{Comp}    \\\midrule
Relevance & $95.71\pm6.22$ & $89.28\pm12.79$ & $81.42\pm14.06$ & $81.42\pm12.45$ & $99.28\pm2.57$ & $92.14\pm11.45$ \\\midrule
Consistency & $95.0\pm5.00$ & $81.42\pm10.59$ & $85.71\pm9.03$ & $81.42\pm11.24$ & $94.28\pm4.94$ & $82.85\pm8.80$ \\\bottomrule
\end{tabular}
\label{tab:error_bar_substudy}
\end{table*}

\section{Fuser Scheduler}

\begin{figure}
    \centering
    \includegraphics[width=\linewidth]{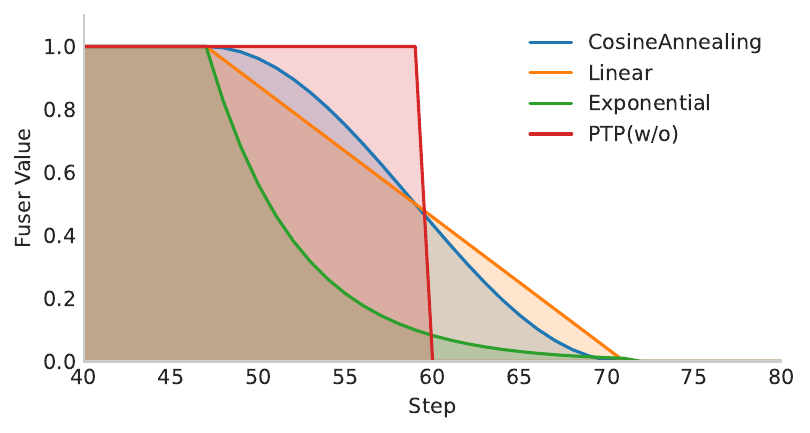}
    \caption{\textbf{Decay lines of Fuser with different schedulers.} We assume the cross replace steps is 0.6 and the decay window is 12.}
    \label{fig:curve}
\end{figure}

We utilize schedulers to help mix the attention maps of the source audio and the edited audio during the diffusion steps. Specifically, through ablation studies, we find that the CosineAnnealing scheduler is particularly effective in aiding the editing process. Below, we list all the baseline schedulers considered. All of the listed schedulers start from a higher value $\alpha_{start}$ and decay to a lower value $\alpha_{end}$ here, from a start timestep $t_0$ in a given window size $t_w$: 
\label{sec:sub:sche}
\subsection{Binaray}
The scheduler starts with a high value $\alpha_{start}$ for the first  $t_0$ steps. After that, it returns $\alpha_{end}$. Leveraging this scheduler converts the \textit{Fuser} into a similar way in \ac{ptp} when mixing attention maps.
\begin{equation}
f(t) = 
\begin{cases} 
\alpha_{start} & \text{if } t < t_0 \\
\alpha_{end} & \text{if } t \geq t_0 
\end{cases}
\end{equation}
\subsection{Linear}
Linear scheduler implements a linear decay of the value from $\alpha_{start}$ to $\alpha_{end}$ over a specified number of steps $t_w$.
\begin{equation}
f(t) = 
\begin{cases}
\alpha_{start} & \text{if } t < t_0 \\
\alpha_{start} - \frac{(t - t_0)}{t_w} \cdot (\alpha_{start} - \alpha_{end}) & \text{if } t_0 \leq t \leq t_0+t_w  \\
\alpha_{end} & \text{if } t \geq t_0+t_w\\
\end{cases}
\end{equation}
\subsection{Exponential}
The exponential scheduler implements an exponential decay of the value from $\alpha_{start}$ to $\alpha_{end}$ over a specified number of steps $t_w$, with the decay rate $r_{decay}$:
\begin{equation}
r_{decay} = (\frac{\alpha_{end}}{\alpha_{start}}) ^ {(\frac{1}{t_w})}
\end{equation}
\begin{equation}
f(t) = 
\begin{cases}
\alpha_{start} & \text{if } t < t_0 \\
\alpha_{start} \times r_{decay}^{t - t_0} & \text{if } t_0 \leq t \leq t_0+t_w \\
\alpha_{end} & \text{if } t \geq t_0+t_w \\
\end{cases}
\end{equation}
\subsection{CosineAnnealing}
The implementation of our CosineAnnealing scheduler mainly uses PyTorch's Cosine Annealing learning rate scheduler, in which the decay starts from $t_0$ to $t_w$.
\begin{equation}
f(t) = 
\begin{cases}
\alpha_{start} & \text{if } t < t_0 \\
\frac{\alpha_{start}}{2} \times \left( \cos\left(\frac{\pi \times (t - t_0)}{t_w}\right) + 1 \right) & \text{if } t_0 \leq t \leq t_0+t_w \\
\alpha_{end} & \text{if } t \geq t_0+t_w \\
\end{cases}
\end{equation}

\section{Test Set Construction}
\label{sec:supp:testset}
For instance, we choose an audio clip \( a_1 \) with the prompt \( p_1 = \text{"a cat meowing"} \) and another \( a_2 \) with \( p_2 = \text{"a baby crying"} \). We combine these as the source audio, generating its description using GPT \cite{openai2023gpt}. For Audio Editing, we select a third audio clip \( a_3 \) with \( p_3 = \text{"a dog barking"} \). \( a_3 \) is randomly replaced with one of the earlier audio clips and merged with the remaining clip to form the target audio for assessment. A corresponding target description is also generated. The data format for this scenario is denoted as \( a_1 + a_2 \xrightarrow{} a_1 + a_3 \). For the audio refinement task, we consider adding adjective descriptions to one of the original audio pieces and regenerate to get the refined audio \( \widetilde a \), thus \( a_1 + a_2 \xrightarrow{} a_1 + \widetilde a_2 \). In the audio reweighting task, the format is \( a_1 + a_2 \xrightarrow{} a_1 + \alpha \cdot a_2 \), wherein the chosen edit target is reweighted prior to merging with another audio clip.

\section{More Ablation Studies}
\label{sec:supp:abl}
\subsection{Generation Bootstrapping}
\begin{table}[!ht]
\small
\caption{\textbf{The influence of CLAP selection.}}
\centering
\resizebox{\linewidth}{!}{
\begin{tabular}{cccccc}
\toprule
\textbf{Select Groups} & \textbf{FAD}{\color{highlightgreen} $\downarrow$} & \textbf{LSD}{\color{highlightgreen} $\downarrow$} & \textbf{FD}{\color{highlightgreen} $\downarrow$} & \textbf{KL}{\color{highlightgreen} $\downarrow$} & \textbf{CLAP}{\color{highlightred} $\uparrow$} \\ \midrule
 1 (w/ bootstrap) & 3.68 & 1.50 & 28.32 & 1.42 & 0.63 \\\midrule
 3 (w/ bootstrap) & 3.40 & 1.47 & 23.91 & 1.44 & 0.68 \\\midrule
 10 (w/ bootstrap) & 3.38 & 1.47 & 24.24 & 1.32 & 0.70 \\\midrule
 10 (w/o bootstrap)& 5.52 & 2.24 & 52.41 & 1.89 & 0.68 \\ \bottomrule
\end{tabular}}
\label{tab:regen_clap}
\end{table}
Bootstrapping generation configurations plays a crucial role in ensuring effective editing, as different editing tasks involve varying components of the original audio, making it challenging to determine the extent of guidance required for the editing process. While our work is the first to explore this issue in editing tasks, similar selection methods using CLAP have been employed in generation tasks to enhance effectiveness. Generally, existing methods create batches of audio and employ CLAP to select the best one. For instance, AudioLDM2~\cite{liu2023audioldm2} use CLAP to filter generated audios, which they call "clap filtering." Similarly, Audiobox~\cite{vyas2023audiobox} utilizes CLAP reranking with $N = 8$ or even $16$ samples using the sound clap model.

We conducted further studies, as illustrated in \cref{tab:regen_clap}. First, we attempted to combine batch sampling methods with sample numbers ranging from 1 to 10. The results indicate that this approach only yields a slight improvement in metrics. Next, we removed the bootstrapping module, allowing our model to generate batches of edited audio with a fixed guidance scale of 3 (consistent with the original Tango model). This scenario resulted in a significant performance decline, highlighting the importance of guidance bootstrapping within the entire editing pipeline.

\subsection{Regeneration Steps}
We observe that regenerating the audio according to the inversion guidance aids in the recovery of the original audio content and structure. However, it negatively affects the audio quality and fidelity, as a more extensive mixture of the attention map results in increased fusion during the intermediate step. This could explain some of the low metrics observed for the \ac{method} in \cref{sec:results}. To address this issue, intuitively we can incorporate extra diffusion steps, as previously discussed in many previous paper. However, it is crucial to acknowledge the inherent trade-off: while additional diffusion steps lack inversion guidance, increasing their number may cause the generated audio to lose more information from the source audio, as shown in \cref{tab:regen_step}.

\begin{table}[!ht]
\small
\caption{\textbf{The influence of re-diffusion.}}
\centering
\resizebox{\linewidth}{!}{
\begin{tabular}{cccccc}
\toprule
\textbf{Extra Steps} & \textbf{FAD}{\color{highlightgreen} $\downarrow$} & \textbf{LSD}{\color{highlightgreen} $\downarrow$} & \textbf{FD}{\color{highlightgreen} $\downarrow$} & \textbf{KL}{\color{highlightgreen} $\downarrow$} & \textbf{CLAP}{\color{highlightred} $\uparrow$} \\ \midrule
 0 & 3.72 & 1.68 & 29.12 & 1.51 & 0.76 \\\midrule
 10 & 12.43 & 3.39 & 84.90 &  1.96 & 0.82 \\\midrule
 20 & 14.95 & 3.35 & 77.55 & 1.62 & 0.74 \\\midrule
 50 & 12.80 & 3.35 & 71.40 & 2.03 & 0.78 \\ \bottomrule
\end{tabular}}
\label{tab:regen_step}
\end{table}

\section{Limitations}
Our proposed audio editing method relies on accurate inversion of the given audio. If the audio content falls outside the model's trained domain, precise editing becomes challenging. Furthermore, our approach primarily focuses on modifications at the attention map level, which inherently restricts the extent of the edits. It may not be suitable for more substantial structural alterations.

%% file: supp_fig_replace.tex
\begin{figure}[!ht]
\centering
    \begin{subfigure}[b]{\linewidth}
    \centering
    \includegraphics[width=\linewidth]{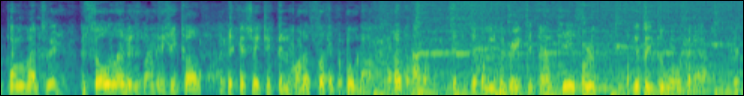}
    \caption{A \textcolor{red}{man} talking and a soft music.wav (Source)}
    \end{subfigure}
    \begin{subfigure}[b]{\linewidth}
    \centering
    \includegraphics[width=\linewidth]{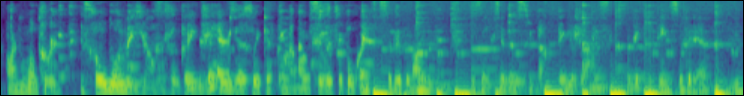}
    \caption{A \textcolor{red}{woman} talking and a soft music.wav (Edited by PPAE)}
    \end{subfigure}
    \begin{subfigure}[b]{\linewidth}
    \centering
    \includegraphics[width=\linewidth]{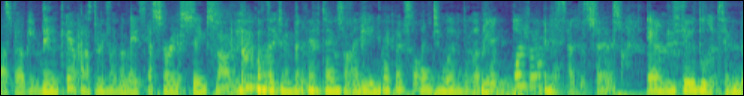}
    \caption{A \textcolor{red}{woman} talking and a soft music.wav (Regenerated)}
    \end{subfigure}
\caption{\textbf{Case Study (Audio Replace)}}
\end{figure}

\begin{figure}[!ht]
\centering
    \begin{subfigure}[b]{\linewidth}
    \centering
    \includegraphics[width=\linewidth]{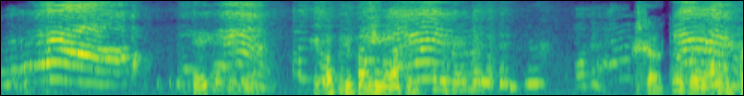}
    \caption{A \textcolor{red}{man} talking and a baby crying.wav (Source)}
    \end{subfigure}
    \begin{subfigure}[b]{\linewidth}
    \centering
    \includegraphics[width=\linewidth]{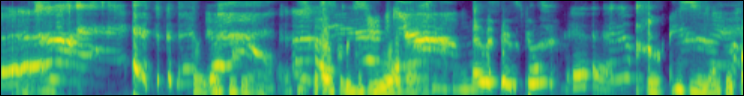}
    \caption{A \textcolor{red}{woman} talking and a baby crying.wav (Edited by PPAE)}
    \end{subfigure}
    \begin{subfigure}[b]{\linewidth}
    \centering
    \includegraphics[width=\linewidth]{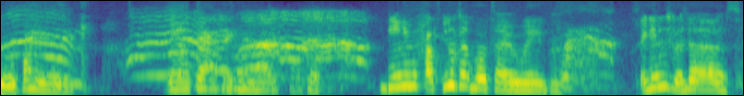}
    \caption{A \textcolor{red}{woman} talking and a baby crying.wav (Regenerated)}
    \end{subfigure}
\caption{\textbf{Case Study (Audio Replace)}}
\end{figure}

\begin{figure}[!ht]
\centering
    \begin{subfigure}[b]{\linewidth}
    \centering
    \includegraphics[width=\linewidth]{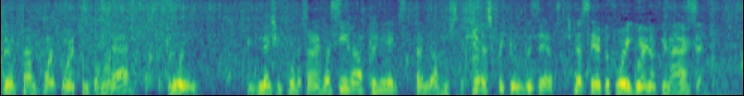}
    \caption{The water flowing and a \textcolor{red}{man} talking.wav (Source)}
    \end{subfigure}
    \begin{subfigure}[b]{\linewidth}
    \centering
    \includegraphics[width=\linewidth]{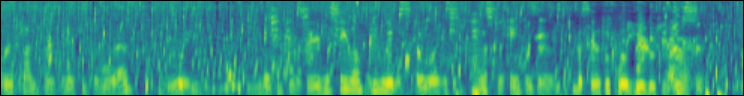}
    \caption{The water flowing and a \textcolor{red}{woman} talking.wav (Edited by PPAE)}
    \end{subfigure}
    \begin{subfigure}[b]{\linewidth}
    \centering
    \includegraphics[width=\linewidth]{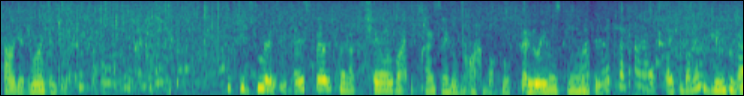}
    \caption{The water flowing and a \textcolor{red}{woman} talking.wav (Regenerated)}
    \end{subfigure}
\caption{\textbf{Case Study (Audio Replace)}}
\end{figure}

\begin{figure}[!ht]
\centering
    \begin{subfigure}[b]{\linewidth}
    \centering
    \includegraphics[width=\linewidth]{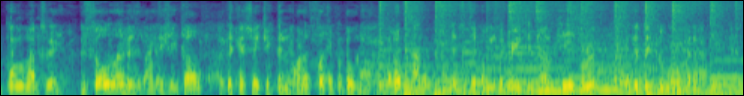}
    \caption{A man talking and a \textcolor{red}{jazz} music.wav (Source)}
    \end{subfigure}
    \begin{subfigure}[b]{\linewidth}
    \centering
    \includegraphics[width=\linewidth]{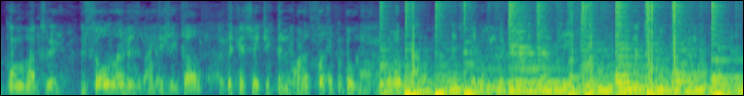}
    \caption{A man talking and a \textcolor{red}{rock} music.wav (Edited by PPAE)}
    \end{subfigure}
    \begin{subfigure}[b]{\linewidth}
    \centering
    \includegraphics[width=\linewidth]{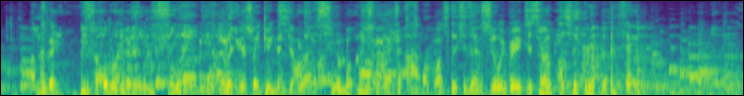}
    \caption{A man talking and a \textcolor{red}{rock} music.wav (Regenerated)}
    \end{subfigure}
\caption{\textbf{Case Study (Audio Replace)}}
\end{figure}

\begin{figure}[!ht]
\centering
    \begin{subfigure}[b]{\linewidth}
    \centering
    \includegraphics[width=\linewidth]{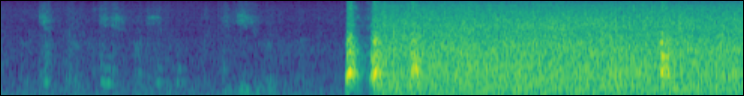}
    \caption{A dog barking and \textcolor{red}{water flowing}.wav (Source)}
    \end{subfigure}
    \begin{subfigure}[b]{\linewidth}
    \centering
    \includegraphics[width=\linewidth]{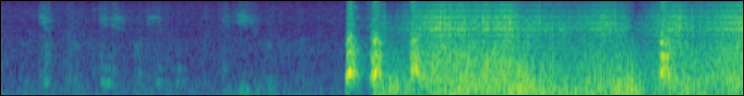}
    \caption{A dog barking and \textcolor{red}{wind blowing}.wav (Edited by PPAE)}
    \end{subfigure}
    \begin{subfigure}[b]{\linewidth}
    \centering
    \includegraphics[width=\linewidth]{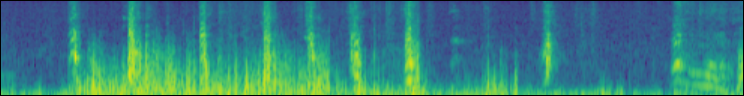}
    \caption{A dog barking and \textcolor{red}{wind blowing}.wav (Regenerated)}
    \end{subfigure}
\caption{\textbf{Case Study (Audio Replace)}}
\end{figure}

\begin{figure}[!ht]
\centering
    \begin{subfigure}[b]{\linewidth}
    \centering
    \includegraphics[width=\linewidth]{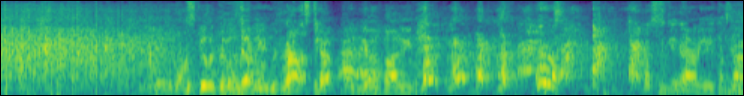}
    \caption{a man applause and a \textcolor{red}{dog barking}.wav (Source)}
    \end{subfigure}
    \begin{subfigure}[b]{\linewidth}
    \centering
    \includegraphics[width=\linewidth]{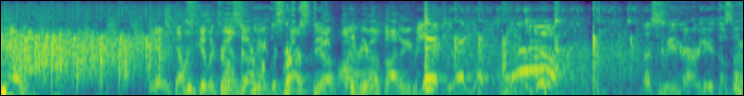}
    \caption{a man applause and a \textcolor{red}{metal collision}.wav (Edited by PPAE)}
    \end{subfigure}
    \begin{subfigure}[b]{\linewidth}
    \centering
    \includegraphics[width=\linewidth]{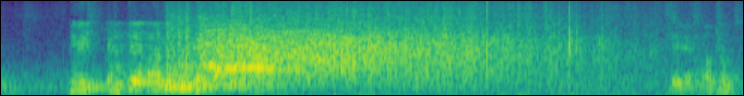}
    \caption{a man applause and a \textcolor{red}{metal collision}.wav (Regenerated)}
    \end{subfigure}
\caption{\textbf{Case Study (Audio Replace)}}
\end{figure}

%% file: supp_fig_refine.tex
\begin{figure}[!htp]
\centering
    \begin{subfigure}[b]{\linewidth}
    \centering
    \includegraphics[width=\linewidth]{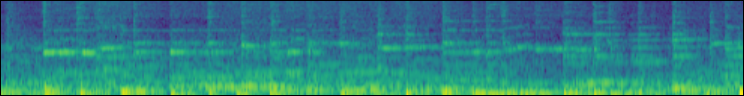}
    \caption{A piece of music.wav (Source)}
    \end{subfigure}
    \begin{subfigure}[b]{\linewidth}
    \centering
    \includegraphics[width=\linewidth]{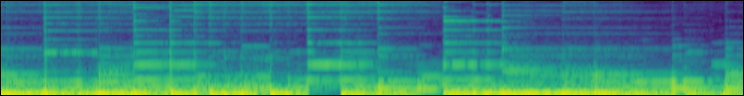}
    \caption{A piece of \textcolor{red}{rock} music.wav (Edited by PPAE)}
    \end{subfigure}
    \begin{subfigure}[b]{\linewidth}
    \centering
    \includegraphics[width=\linewidth]{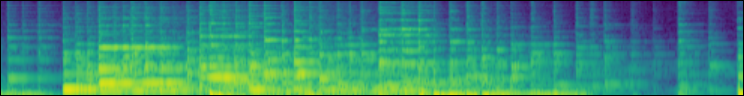}
    \caption{A piece of \textcolor{red}{rock} music.wav (Regenerated)}
    \end{subfigure}
\caption{\textbf{Case Study (Audio Refine)}}
\end{figure}

\begin{figure}[!htp]
\centering
    \begin{subfigure}[b]{\linewidth}
    \centering
    \includegraphics[width=\linewidth]{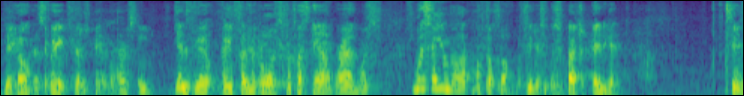}
    \caption{A man talking.wav (Source)}
    \end{subfigure}
    \begin{subfigure}[b]{\linewidth}
    \centering
    \includegraphics[width=\linewidth]{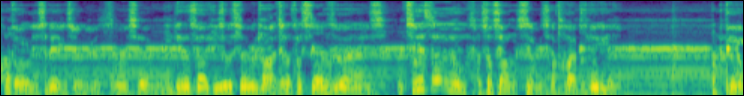}
    \caption{A man talking \textcolor{red}{in a large room}.wav (Edited by PPAE)}
    \end{subfigure}
    \begin{subfigure}[b]{\linewidth}
    \centering
    \includegraphics[width=\linewidth]{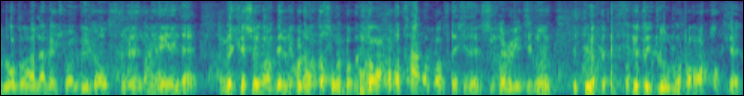}
    \caption{A man talking \textcolor{red}{in a large room}.wav (Regenerated)}
    \end{subfigure}
    
\caption{\textbf{Case Study (Audio Refine)}}
\end{figure}

\begin{figure}[!htp]
\centering
    \begin{subfigure}[b]{\linewidth}
    \centering
    \includegraphics[width=\linewidth]{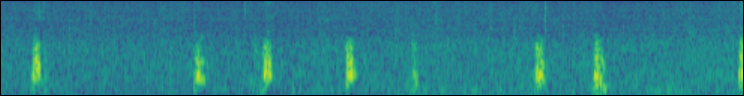}
    \caption{A dog barking and raining.wav (Source)}
    \end{subfigure}
    \begin{subfigure}[b]{\linewidth}
    \centering
    \includegraphics[width=\linewidth]{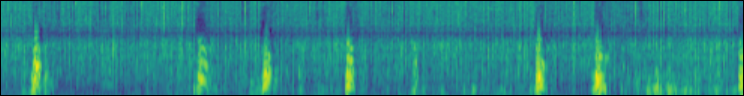}
    \caption{A dog barking and raining \textcolor{red}{heavily}.wav (Edited by PPAE)}
    \end{subfigure}
    \begin{subfigure}[b]{\linewidth}
    \centering
    \includegraphics[width=\linewidth]{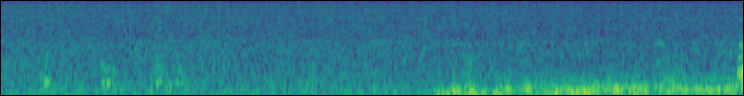}
    \caption{A dog barking and raining \textcolor{red}{heavily}.wav (Regenerated)}
    \end{subfigure}
\caption{\textbf{Case Study (Audio Refine)}}
\end{figure}

\begin{figure}[!htp]
\centering
    \begin{subfigure}[b]{\linewidth}
    \centering
    \includegraphics[width=\linewidth]{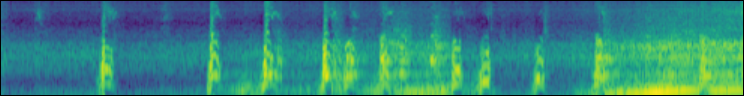}
    \caption{A dog barking and a car engine running.wav (Source)}
    \end{subfigure}
    \begin{subfigure}[b]{\linewidth}
    \centering
    \includegraphics[width=\linewidth]{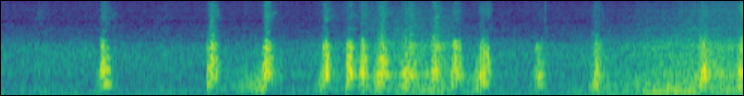}
    \caption{A dog barking and a car engine running \textcolor{red}{in a far place}.wav (Edited by PPAE)}
    \end{subfigure}
    \begin{subfigure}[b]{\linewidth}
    \centering
    \includegraphics[width=\linewidth]{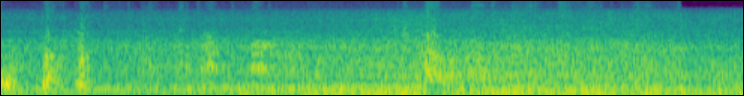}
    \caption{A dog barking and a car engine running \textcolor{red}{in a far place}.wav (Regenerated)}
    \end{subfigure}
\caption{\textbf{Case Study (Audio Refine)}}
\end{figure}

\begin{figure}[!htp]
\centering
    \begin{subfigure}[b]{\linewidth}
    \centering
    \includegraphics[width=\linewidth]{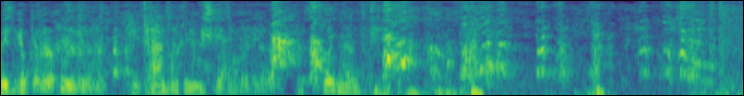}
    \caption{A woman talking and a dog barking.wav (Source)}
    \end{subfigure}
    \begin{subfigure}[b]{\linewidth}
    \centering
    \includegraphics[width=\linewidth]{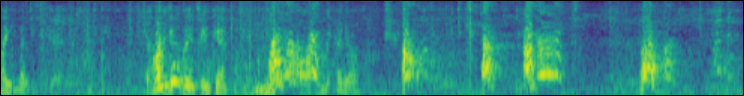}
    \caption{A woman talking and a dog barking \textcolor{red}{loudly}.wav (Edited by PPAE)}
    \end{subfigure}
    \begin{subfigure}[b]{\linewidth}
    \centering
    \includegraphics[width=\linewidth]{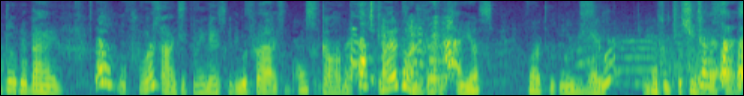}
    \caption{A woman talking and a dog barking \textcolor{red}{loudly}.wav (Regenerated)}
    \end{subfigure}
\caption{\textbf{Case Study (Audio Refine)}}
\end{figure}

\begin{figure}[!htp]
\centering
    \begin{subfigure}[b]{\linewidth}
    \centering
    \includegraphics[width=\linewidth]{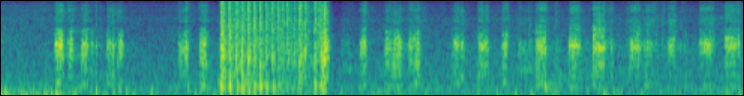}
    \caption{A duck quacking and gunshot.wav (Source)}
    \end{subfigure}
    \begin{subfigure}[b]{\linewidth}
    \centering
    \includegraphics[width=\linewidth]{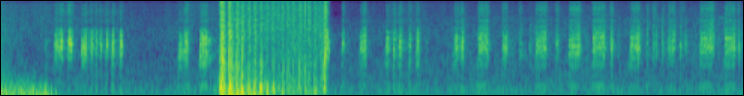}
    \caption{A duck quacking and \textcolor{red}{loud} gunshot.wav (Edited by PPAE)}
    \end{subfigure}
    \begin{subfigure}[b]{\linewidth}
    \centering
    \includegraphics[width=\linewidth]{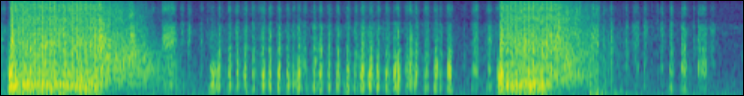}
    \caption{A duck quacking and \textcolor{red}{loud} gunshot.wav (Regenerated)}
    \end{subfigure}
\caption{\textbf{Case Study (Audio Refine)}}
\end{figure}

%% file: supp_fig_reweight.tex
\begin{figure}[!ht]
\centering
    \begin{subfigure}[b]{\linewidth}
    \centering
    \includegraphics[width=\linewidth]{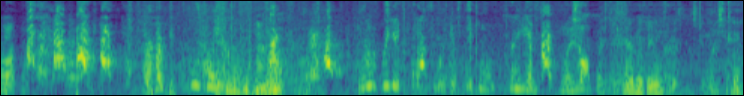}
    \caption{Someone talking and a dog \textcolor{red}{barking}.wav, $c$ = 2}
    \end{subfigure}
    \begin{subfigure}[b]{\linewidth}
    \centering
    \includegraphics[width=\linewidth]{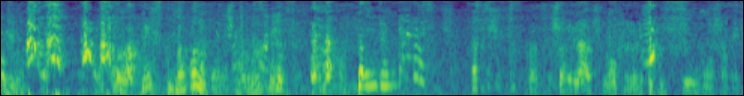}
    \caption{Someone talking and a dog \textcolor{red}{barking}.wav, $c$ = 0}
    \end{subfigure}
    \begin{subfigure}[b]{\linewidth}
    \centering
    \includegraphics[width=\linewidth]{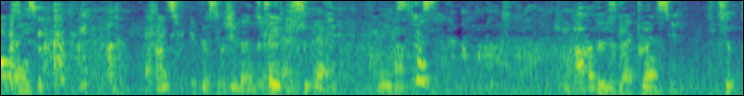}
    \caption{Someone talking and a dog \textcolor{red}{barking}.wav, $c$ = -2}
    \end{subfigure}
\caption{\textbf{Case Study (Audio Reweight)}}
\end{figure}

\begin{figure}[!ht]
\centering
    \begin{subfigure}[b]{\linewidth}
    \centering
    \includegraphics[width=\linewidth]{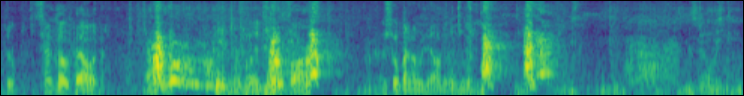}
    \caption{A man talking and a dog \textcolor{red}{barking}.wav, $c$ = 2}
    \end{subfigure}
    \begin{subfigure}[b]{\linewidth}
    \centering
    \includegraphics[width=\linewidth]{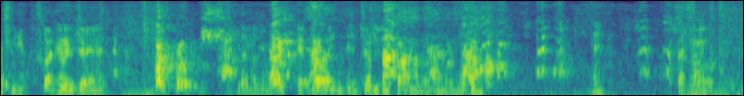}
    \caption{A man talking and a dog \textcolor{red}{barking}.wav, $c$ = 0}
    \end{subfigure}
    \begin{subfigure}[b]{\linewidth}
    \centering
    \includegraphics[width=\linewidth]{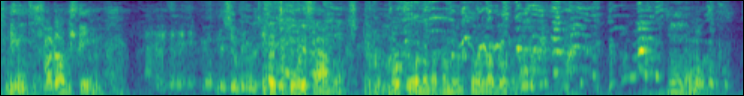}
    \caption{A man talking and a dog \textcolor{red}{barking}.wav, $c$ = -2}
    \end{subfigure}
\caption{\textbf{Case Study (Audio Reweight)}}
\end{figure}

\begin{figure}[!ht]
\centering
    \begin{subfigure}[b]{\linewidth}
    \centering
    \includegraphics[width=\linewidth]{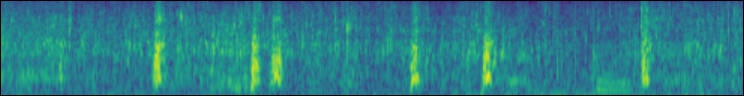}
    \caption{The water flowing and a dog \textcolor{red}{barking}.wav, $c$ = 2}
    \end{subfigure}
    \begin{subfigure}[b]{\linewidth}
    \centering
    \includegraphics[width=\linewidth]{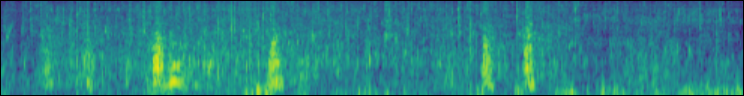}
    \caption{The water flowing and a dog \textcolor{red}{barking}.wav, $c$ = 0}
    \end{subfigure}
    \begin{subfigure}[b]{\linewidth}
    \centering
    \includegraphics[width=\linewidth]{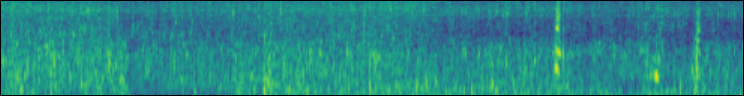}
    \caption{The water flowing and a dog \textcolor{red}{barking}.wav, $c$ = -2}
    \end{subfigure}
\caption{\textbf{Case Study (Audio Reweight)}}
\end{figure}

\begin{figure}[!ht]
\centering
    \begin{subfigure}[b]{\linewidth}
    \centering
    \includegraphics[width=\linewidth]{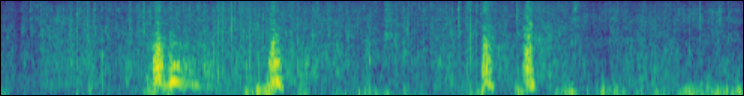}
    \caption{The rain falling and a dog \textcolor{red}{barking}.wav, $c$ = 2}
    \end{subfigure}
    \begin{subfigure}[b]{\linewidth}
    \centering
    \includegraphics[width=\linewidth]{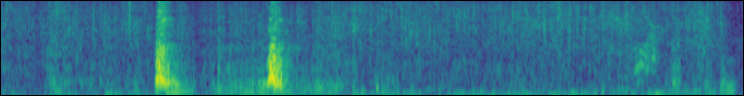}
    \caption{The rain falling and a dog \textcolor{red}{barking}.wav, $c$ = 0}
    \end{subfigure}
    \begin{subfigure}[b]{\linewidth}
    \centering
    \includegraphics[width=\linewidth]{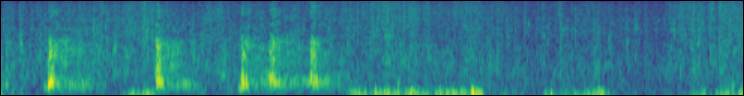}
    \caption{The rain falling and a dog \textcolor{red}{barking}.wav, $c$ = -2}
    \end{subfigure}
\caption{\textbf{Case Study (Audio Reweight)}}
\end{figure}

\begin{figure}[!ht]
\centering
    \begin{subfigure}[b]{\linewidth}
    \centering
    \includegraphics[width=\linewidth]{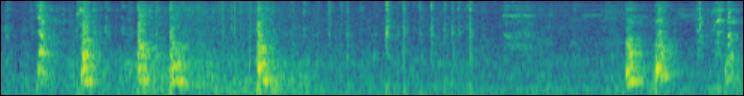}
    \caption{The machine clicks and a dog \textcolor{red}{barking}.wav, $c$ = 2}
    \end{subfigure}
    \begin{subfigure}[b]{\linewidth}
    \centering
    \includegraphics[width=\linewidth]{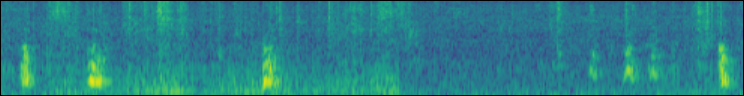}
    \caption{The machine clicks and a dog \textcolor{red}{barking}.wav, $c$ = 0}
    \end{subfigure}
    \begin{subfigure}[b]{\linewidth}
    \centering
    \includegraphics[width=\linewidth]{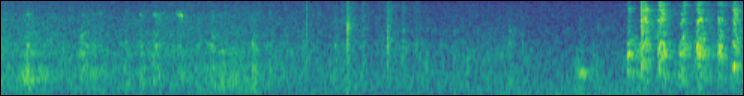}
    \caption{The machine clicks and a dog \textcolor{red}{barking}.wav, $c$ = -2}
    \end{subfigure}
\caption{\textbf{Case Study (Audio Reweight)}}
\end{figure}

\begin{figure}[!ht]
\centering
    \begin{subfigure}[b]{\linewidth}
    \centering
    \includegraphics[width=\linewidth]{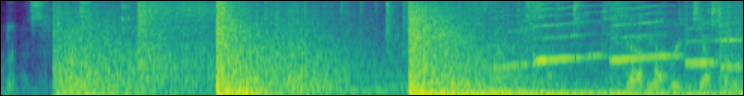}
    \caption{A man talking and \textcolor{red}{firework}.wav, $c$ = 2}
    \end{subfigure}
    \begin{subfigure}[b]{\linewidth}
    \centering
    \includegraphics[width=\linewidth]{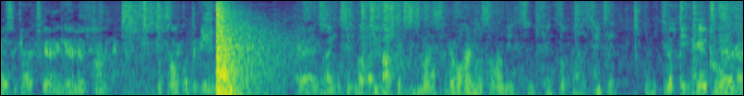}
    \caption{A man talking and \textcolor{red}{firework}.wav, $c$ = 0}
    \end{subfigure}
    \begin{subfigure}[b]{\linewidth}
    \centering
    \includegraphics[width=\linewidth]{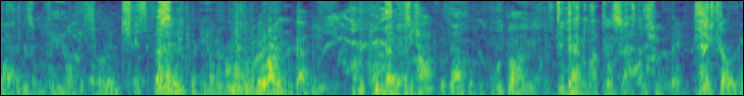}
    \caption{A man talking and \textcolor{red}{firework}.wav, $c$ = -2}
    \end{subfigure}
\caption{\textbf{Case Study (Audio Reweight)}}
\end{figure}